\documentclass[fleqn,10pt]{wlscirep}
\usepackage[utf8]{inputenc}
\usepackage[T1]{fontenc}
\usepackage{graphicx}
\usepackage{booktabs}
\usepackage{subcaption}
\usepackage[english]{babel}
\usepackage{pdflscape}
\usepackage{multirow}
\usepackage{hyperref}
\usepackage{xurl}
\hypersetup{
    bookmarksopen=true,        
    bookmarksopenlevel=1,      
    colorlinks=true,
    pdfstartview=FitH,
    pdfpagemode=UseOutlines    
}
\usepackage{dsfont}
\usepackage{indentfirst}      
\usepackage{lmodern}

\title{Noise-adaptive hybrid quantum convolutional neural networks based on depth-stratified feature extraction}

\author[1,*]{Taehyun Kim}
\author[1,2]{Israel F. Araujo}
\author[1,3,4,+]{Daniel K. Park}

\affil[1]{Department of Statistics and Data Science, Yonsei University, Seoul, 03722, Republic of Korea}
\affil[2]{data cybernetics ssc GmbH, Landsberg am Lech 86899, Germany}
\affil[3]{Department of Applied Statistics, Yonsei University, Seoul 03722, Republic of Korea}
\affil[4]{Department of Quantum Information, Yonsei University, Seoul 03722, Republic of Korea}
\affil[*]{taehyunkim@yonsei.ac.kr}
\affil[+]{dkd.park@yonsei.ac.kr}

\begin{abstract}
Hierarchical quantum classifiers, such as quantum convolutional neural networks (QCNNs), represent recent progress toward designing effective and feasible architectures for quantum classification. However, their performance on near-term quantum hardware remains highly sensitive to noise accumulation across circuit depth, calling for strategies beyond circuit-architecture design alone. We propose a noise-adaptive hybrid QCNN that improves classification under noise by exploiting depth-stratified intermediate measurements. Instead of discarding qubits removed during pooling operations, we measure them and use the resulting outcomes as classical features that are jointly processed by a classical neural network. This hybrid hierarchical design enables noise-adaptive inference by integrating quantum intermediate measurements with classical post-processing. Systematic experiments across multiple circuit sizes and noise settings, including hardware-calibrated noise models derived from IBM Quantum backend data, demonstrate more stable convergence, reduced loss variability, and consistently higher classification accuracy compared with standard QCNNs. Moreover, we observe that this performance advantage significantly amplifies as the circuit size increases, confirming that the hybrid architecture mitigates the scaling limitations of standard architectures. Notably, the multi-basis measurement variant attains performance close to the noiseless limit even under realistic noise. While demonstrated for QCNNs, the proposed depth-stratified feature extraction applies more broadly to hierarchical quantum classifiers that progressively discard qubits.
\end{abstract}

\begin{document}

\pdfbookmark[0]{Noise-adaptive hybrid quantum convolutional neural networks based on depth-stratified feature extraction}{title}
\flushbottom
\maketitle

\thispagestyle{empty}





\section{Introduction}

Quantum machine learning (QML) studies how quantum information processing techniques can be utilized in learning tasks, with the aim of developing new learning frameworks that go beyond purely classical approaches~\cite{biamonte2017quantum, schuld2019quantum}. While a variety of potential advantages of QML have been explored in theory~\cite{lloyd2014quantum, rebentrost2014quantum, cross2015quantum, harrow2017quantum}, its practical relevance increasingly depends on approaches that can be implemented on noisy intermediate-scale quantum (NISQ) devices~\cite{preskill2018quantum} and early fault-tolerant quantum processors. Motivated by this consideration, a broad class of QML approaches has emerged, including variational quantum circuits and hybrid quantum–classical models, which seek to balance expressive modeling capacity with practical implementability under realistic hardware constraints~\cite{cerezo2021variational, benedetti2019parameterized, schuld2020circuit}. Despite these developments, quantum noise remains a fundamental challenge in practice, often limiting the reliability and scalability of learning performance on current quantum hardware.

Within this broader context of QML, an important line of research involves designing structured quantum model architectures that mitigate these constraints while promoting trainability and generalization~\cite{cerezo2021variational, pesah2021absence, banchi2021generalization, jaderberg2024let, kim2025expressivity}. In particular, hierarchical circuit structures have attracted considerable attention due to their ability to efficiently capture complex and long-range correlations using logarithmic circuit depth~\cite{grant2018hierarchical, cong2019quantum, lourens2023hierarchical, zapletal2024error}. These architectures draw inspiration from tensor networks---widely used in quantum many-body physics to represent both quantum states and quantum circuits---and naturally mirror the layered design principles of classical neural networks \cite{orus2014practical, shi2006classical, verstraete2008matrix}. This perspective provides a systematic foundation for constructing quantum models that combine expressive structure with favorable learning properties under realistic hardware constraints.

Early realizations of this hierarchical paradigm were introduced by Grant et al.~\cite{grant2018hierarchical}, who proposed quantum classifiers based on tree-structured tensor-network ans\"{a}tze, including the Tree Tensor Network (TTN)~\cite{shi2006classical} and the Multi-scale Entanglement Renormalization Ansatz (MERA)~\cite{vidal2008class}. These models implement classification through successive layers of local unitaries that progressively reduce the number of active qubits, yielding a compact final readout.

Building on the same hierarchical principles, Cong et al.~\cite{cong2019quantum} introduced the Quantum Convolutional Neural Network (QCNN), which adapts the inductive bias of classical convolutional neural networks to the quantum setting. The QCNN extracts features using local entangling gates to capture spatial correlations, followed by pooling operations that reduce the effective dimensionality of the quantum state. By imposing translational invariance through shared unitary parameters within each layer, the QCNN improves parameter efficiency and scalability. The QCNN has demonstrated strong performance in quantum phase recognition~\cite{cong2019quantum, hur2024understanding} and classical image classification tasks~\cite{hur2022quantum}, while maintaining a logarithmic circuit depth that alleviates barren plateau issues~\cite{pesah2021absence}.

Despite these benefits, robustness to noise remains a bottleneck for deployment on near-term quantum devices~\cite{preskill2018quantum}. In standard QCNN architectures, classification decisions are based solely on the final remaining qubit after successive pooling operations, while qubits discarded during pooling---commonly referred to as trash qubits---are ignored. In ideal, noiseless settings, this design choice is well justified. However, in the presence of noise, qubits discarded at earlier stages of the circuit are less exposed to cumulative gate errors and decoherence, and may therefore retain higher-fidelity information about the encoded input data. Leveraging this intermediate information provides an opportunity to enhance the noise robustness of hierarchical quantum classifiers.

Prior work on noise robustness in quantum neural networks has largely relied on error mitigation techniques for final measurements, domain-specific circuit constraints, or comparative noise resilience benchmarking. For instance, Lee et al. \cite{lee2023scalable} utilized classical neural networks to mitigate readout measurement errors, whereas our approach targets noise accumulation across the circuit depth. Zapletal et al. \cite{zapletal2024error} improved noise resilience of QCNN by restricting the model to constant-depth circuits with Boolean post-processing, a method tailored for specific topological phases rather than general-purpose classification. In parallel, Ahmed et al. \cite{ahmed2025quantum} provided an extensive comparative benchmark of distinct hybrid architectures---such as QCNNs, Quanvolutional Neural Networks, and Quantum Transfer Learning---under various noise channels. However, their study focused on evaluating the intrinsic resilience of existing models rather than introducing architectural modifications to actively recover lost information from discarded trash qubits.

Motivated by these observations and the previously overlooked potential of trash qubits, we propose a noise-adaptive hybrid quantum–classical hierarchical model that explicitly leverages information encoded in trash qubits via depth-stratified feature extraction. Instead of discarding these qubits, we measure them at each pooling stage and feed the resulting outcomes into a classical neural network to form the final prediction. The quantum circuit and the classical neural network are jointly trained end to end, enabling the classical model to adaptively combine information collected at different circuit depths, which may be affected by noise to varying degrees. We consider two measurement settings: measurements in the computational basis ($Z$) and multi-basis measurements using ${X,Y,Z}$, allowing us to investigate whether richer representations of the quantum states further enhance noise robustness and model expressivity under diverse noise conditions.

To evaluate the effectiveness of this depth-stratified feature extraction strategy, we benchmark our approach against the standard QCNN model~\cite{cong2019quantum, hur2022quantum, lee2025optimizing}. We evaluate two hybrid quantum–classical variants---QCNNs augmented with depth-stratified mid-circuit measurements using $Z$ and $\{X,Y,Z\}$ bases---under multiple noise configurations derived from real IBM Quantum backend calibration data. As shown in the following sections, the proposed framework yields substantial and consistent improvements in robustness to noise relative to the standard QCNN. Our analysis further reveals that this advantage amplifies as the circuit scales, effectively preventing the performance collapse observed in the baseline model at larger system sizes. To further clarify the origin of these gains, we complement the performance evaluation with an interpretability analysis that quantifies the contribution of depth-stratified measurements from different circuit layers.

\begin{figure}[htbp]
    \centering
    \includegraphics[width=0.98\textwidth]{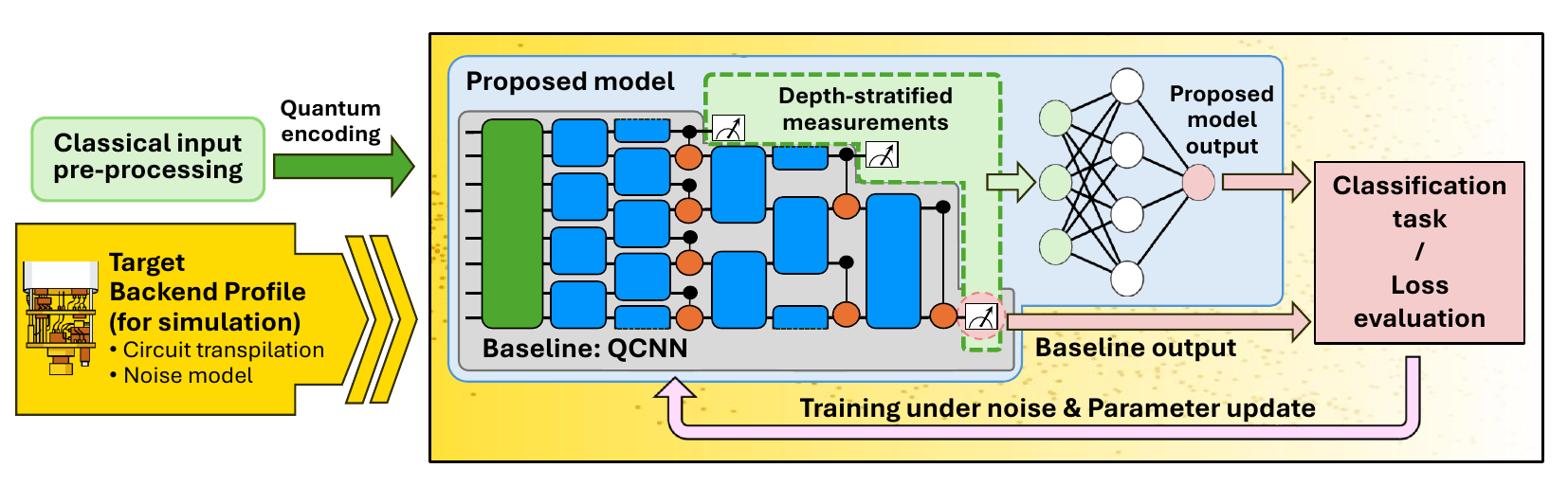}
    \caption{Schematic overview of the proposed noise-robust framework and evaluation pipeline. The workflow initiates with classical input preprocessing (PCA) and quantum encoding (angle encoding) to map data into the quantum state. To rigorously evaluate model performance, a target backend profile is configured to emulate realistic hardware conditions using real device calibration data. The central diagram illustrates the hierarchical architecture: the inner gray block represents the baseline architecture (QCNN), which outputs predictions solely from the final qubit readout, while the enclosing blue block indicates the proposed model (HQCNN-EZ and HQCNN-EM). The proposed architecture recovers information from discarded trash qubits via depth-stratified measurements, processing these outcomes through a classical neural network to compute the output. The surrounding yellow rectangular region delineates the noise-adaptive training loop, where the circuit is transpiled and executed under noise to compute the output and evaluate the loss, enabling the joint update of model parameters directly under noisy execution conditions.}
    \label{fig:hqcnn_schematic_overview}
\end{figure}

\section{Results}\label{sec:Results}
We evaluate the proposed hybrid QCNN framework on supervised learning tasks with the primary goal of characterizing robustness to quantum noise. To this end, numerical simulations are conducted under multiple noise configurations derived from IBM Quantum backend calibration data.

Figure~\ref{fig:hqcnn_schematic_overview} summarizes the noise-adaptive simulation workflow, incorporating device-specific circuit transpilation and an end-to-end training loop under noise. As a baseline, we consider the standard QCNN architecture~\cite{cong2019quantum, hur2022quantum, lee2025optimizing}, which derives its prediction solely from the measurements of the final remaining qubit. We compare this baseline with two hybrid variants that further incorporate depth-stratified single-qubit measurements from each set of qubits discarded during pooling: HQCNN-EZ, which measures the expectation value $\langle Z \rangle$ of each discarded qubit and passes an input size of $\lceil \log_2(n) \rceil$ (where $n$ is the number of qubits used in the circuit) to a classical neural network; and HQCNN-EM, which measures the expectation values $\bigl\{ \langle X \rangle, \langle Y \rangle, \langle Z \rangle \bigr\}$ of each discarded qubit, passing an input dimension three times that of the HQCNN-EZ. Detailed specifications of these quantum circuit and classical neural network architectures are provided in Section~\ref{subsec:Model_architecture}.

For empirical benchmarking, we use binary classification of MNIST digits $0$ and $1$~\cite{deng2012mnist}. This simple task is chosen deliberately to enable a controlled and systematic evaluation of noise robustness, rather than to demonstrate performance on classically challenging instances. Each $28\times28$ grayscale image is flattened and reduced using principal component analysis (PCA) to match the number of input qubits (see Section~\ref{subsec:data_description_and_encoding_strategy}). Unless otherwise stated, all results are obtained under the noisy simulation scenarios described in Section~\ref{subsec:Quantum_noisy_simul_env}. Complementary regression experiments are reported in the Supplementary Information.

\subsection{Performance Stability under Noise}
\label{subsec:Performance_Stability}
\begin{figure}[ht]
\centering
\includegraphics[width=0.95\textwidth]{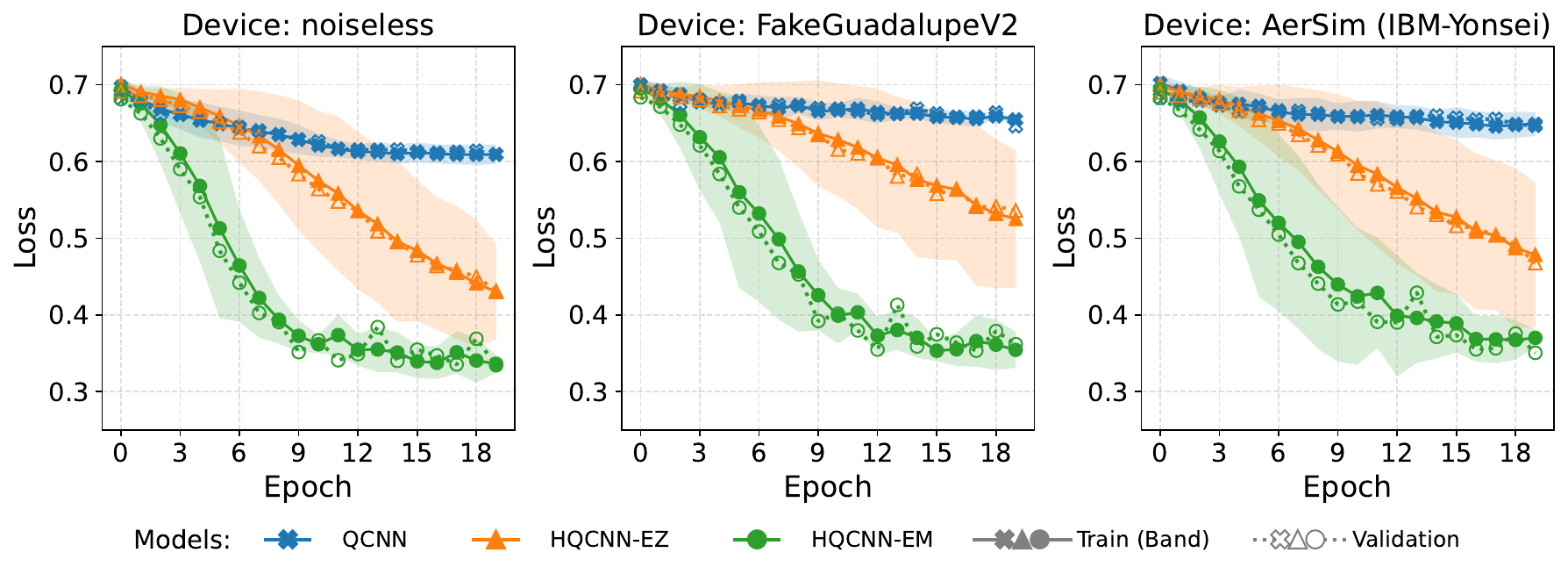}
\caption{Training and validation loss curves for eight-qubit classification models under different noise settings. The plots display the loss trajectories across three simulation environments: noiseless, \texttt{FakeGuadalupeV2}, and \texttt{AerSimulator} configured with the \texttt{IBM\_Yonsei} backend. Solid and dotted lines with filled markers represent the mean of training loss, while dotted lines with open markers indicate the mean validation loss, averaged over five independent iterations. The shaded bands around the training loss curves indicate one standard deviation ($\pm \sigma$) around the mean.}
\label{fig:clf_loss_convergence}
\end{figure}

We begin by examining the training dynamics of each architecture to evaluate whether depth-stratified measurements effectively improve optimization behavior in the presence of noise. Figure~\ref{fig:clf_loss_convergence} illustrates the trajectory of Binary Cross-Entropy (BCE) training (solid) and validation (dotted) losses for the eight-qubit models. The results are reported as the mean across five independent trials with random parameter initialization, with shaded regions indicating one standard deviation. These profiles benchmark the models under three distinct conditions: an ideal noiseless baseline and two realistic noise settings---\texttt{FakeGuadalupeV2} and \texttt{AerSimulator} configured with the \texttt{IBM\_Yonsei} backend---which introduce hardware-specific gate errors and decoherence derived from device calibration data (detailed noise specifications are provided in Supplementary Information).

In this classification task, the baseline QCNN shows limited evidence of effective training, its loss curve displaying only a mild monotonic decrease and converging to higher final values, particularly under noisy conditions. By contrast, both hybrid models, HQCNN-EZ and HQCNN-EM, exhibit a clear and rapid downward trend in the training and validation loss as the number of training epochs increases (Fig.~\ref{fig:clf_loss_convergence}). Notably, when comparing training dynamics, HQCNN-EM consistently exhibits a steeper loss reduction and converges to lower final values, followed by HQCNN-EZ, whereas the baseline QCNN suffers from slower convergence. Furthermore, the training loss band of the HQCNN-EM model is noticeably narrower than that of the baseline QCNN and HQCNN-EZ, demonstrating more stable training. Overall, these results demonstrate that the hybrid architectures possess enhanced trainability and robustness, maintaining high performance across all simulated noisy settings.

To evaluate model performance, we retrieved the optimal model parameters corresponding to the epoch of minimum validation loss. We then computed the BCE loss and classification accuracy on both the training and test datasets (see Section~\ref{subsec:Performance_Metrics} for mathematical formulations). The BCE loss quantifies the divergence between predicted probabilities and ground truth labels, while classification accuracy measures the proportion of correct predictions.

\begin{table}[t]
\centering
\scriptsize 
\setlength{\tabcolsep}{2pt} 
\begin{tabular}{cll|cc|cc|cc}
\toprule
\multicolumn{3}{c|}{\textbf{Classification Task}}                                                                                                                            & \multicolumn{2}{c|}{four qubits}                     & \multicolumn{2}{c|}{eight qubits}                     & \multicolumn{2}{c}{ten qubits}                     \\ \hline
\multicolumn{1}{c|}{Dataset}                & \multicolumn{1}{c|}{Device}                                                                         & \multicolumn{1}{c|}{Model} & BCE   $\downarrow$     & Accuracy (\%) $\uparrow$ & BCE $\downarrow$       & Accuracy (\%) $\uparrow$ & BCE $\downarrow$       & Accuracy (\%) $\uparrow$ \\ \hline
\multicolumn{1}{c|}{\multirow{9}{*}{Train}} & \multicolumn{1}{l|}{\multirow{3}{*}{noiseless}}                                                     & QCNN                       & 0.519 ± 0.034          & \textbf{85.2 ± 1.151}    & 0.610 ± 0.013          & 85.0 ± 2.761             & 0.603 ± 0.033          & 67.8 ± 15.739            \\
\multicolumn{1}{c|}{}                       & \multicolumn{1}{l|}{}                                                                               & HQCNN-EZ                   & 0.443 ± 0.092          & 84.4 ± 0.652             & 0.424 ± 0.064          & \textbf{86.7 ± 1.304}    & 0.343 ± 0.068          & \textbf{88.5 ± 1.904}    \\
\multicolumn{1}{c|}{}                       & \multicolumn{1}{l|}{}                                                                               & HQCNN-EM                   & \textbf{0.379 ± 0.015} & 84.7 ± 1.204             & \textbf{0.324 ± 0.016} & 85.9 ± 0.742             & \textbf{0.327 ± 0.053} & 88.2 ± 0.671             \\ \cline{2-9} 
\multicolumn{1}{c|}{}                       & \multicolumn{1}{l|}{\multirow{3}{*}{FakeGuadalupeV2}}                                               & QCNN                       & 0.562 ± 0.031          & 83.3 ± 2.110             & 0.657 ± 0.005          & 68.2 ± 4.868             & 0.655 ± 0.022          & 58.7 ± 7.547             \\
\multicolumn{1}{c|}{}                       & \multicolumn{1}{l|}{}                                                                               & HQCNN-EZ                   & 0.485 ± 0.104          & 84.2 ± 0.908             & 0.522 ± 0.096          & 83.2 ± 4.508             & 0.482 ± 0.086          & 85.4 ± 2.460             \\
\multicolumn{1}{c|}{}                       & \multicolumn{1}{l|}{}                                                                               & HQCNN-EM                   & \textbf{0.405 ± 0.032} & \textbf{85.5 ± 0.612}    & \textbf{0.362 ± 0.013} & \textbf{85.7 ± 1.095}    & \textbf{0.369 ± 0.053} & \textbf{86.6 ± 2.275}    \\ \cline{2-9} 
\multicolumn{1}{c|}{}                       & \multicolumn{1}{l|}{\multirow{3}{*}{\begin{tabular}[c]{@{}l@{}}AerSim\\ (IBM-Yonsei)\end{tabular}}} & QCNN                       & 0.585 ± 0.026          & 82.8 ± 2.280             & 0.664 ± 0.016          & 66.8 ± 10.091            & 0.651 ± 0.035          & 59.1 ± 2.859             \\
\multicolumn{1}{c|}{}                       & \multicolumn{1}{l|}{}                                                                               & HQCNN-EZ                   & 0.496 ± 0.113          & 81.8 ± 4.396             & 0.489 ± 0.078          & 84.4 ± 2.903             & 0.454 ± 0.084          & 84.7 ± 2.388             \\
\multicolumn{1}{c|}{}                       & \multicolumn{1}{l|}{}                                                                               & HQCNN-EM                   & \textbf{0.403 ± 0.015} & \textbf{85.1 ± 0.652}    & \textbf{0.372 ± 0.038} & \textbf{86.4 ± 1.851}    & \textbf{0.381 ± 0.081} & \textbf{86.7 ± 2.885}    \\ \midrule
\multicolumn{1}{c|}{\multirow{9}{*}{Test}}  & \multicolumn{1}{l|}{\multirow{3}{*}{noiseless}}                                                     & QCNN                       & 0.471 ± 0.048          & 93.0 ± 1.275             & 0.604 ± 0.014          & 88.5 ± 4.123             & 0.597 ± 0.049          & 68.2 ± 17.824            \\
\multicolumn{1}{c|}{}                       & \multicolumn{1}{l|}{}                                                                               & HQCNN-EZ                   & 0.357 ± 0.127          & 92.7 ± 0.570             & 0.390 ± 0.073          & 91.7 ± 1.255             & 0.326 ± 0.073          & 90.9 ± 1.884             \\
\multicolumn{1}{c|}{}                       & \multicolumn{1}{l|}{}                                                                               & HQCNN-EM                   & \textbf{0.253 ± 0.016} & \textbf{93.6 ± 0.652}    & \textbf{0.247 ± 0.013} & \textbf{92.7 ± 0.447}    & \textbf{0.302 ± 0.061} & \textbf{91.2 ± 0.975}    \\ \cline{2-9} 
\multicolumn{1}{c|}{}                       & \multicolumn{1}{l|}{\multirow{3}{*}{FakeGuadalupeV2}}                                               & QCNN                       & 0.521 ± 0.038          & 92.6 ± 1.475             & 0.655 ± 0.007          & 70.1 ± 4.037             & 0.650 ± 0.029          & 59.4 ± 8.771             \\
\multicolumn{1}{c|}{}                       & \multicolumn{1}{l|}{}                                                                               & HQCNN-EZ                   & 0.423 ± 0.141          & 92.0 ± 1.768             & 0.498 ± 0.117          & 85.9 ± 10.359            & 0.458 ± 0.108          & 86.2 ± 5.008             \\
\multicolumn{1}{c|}{}                       & \multicolumn{1}{l|}{}                                                                               & HQCNN-EM                   & \textbf{0.295 ± 0.044} & \textbf{93.2 ± 0.758}    & \textbf{0.279 ± 0.011} & \textbf{92.3 ± 0.758}    & \textbf{0.335 ± 0.078} & \textbf{89.8 ± 2.992}    \\ \cline{2-9} 
\multicolumn{1}{c|}{}                       & \multicolumn{1}{l|}{\multirow{3}{*}{\begin{tabular}[c]{@{}l@{}}AerSim\\ (IBM-Yonsei)\end{tabular}}} & QCNN                       & 0.560 ± 0.031          & 89.8 ± 2.019             & 0.663 ± 0.016          & 66.1 ± 6.950             & 0.644 ± 0.034          & 62.1 ± 5.595             \\
\multicolumn{1}{c|}{}                       & \multicolumn{1}{l|}{}                                                                               & HQCNN-EZ                   & 0.438 ± 0.151          & 90.3 ± 4.396             & 0.455 ± 0.095          & 90.1 ± 3.111             & 0.432 ± 0.109          & 87.6 ± 4.955             \\
\multicolumn{1}{c|}{}                       & \multicolumn{1}{l|}{}                                                                               & HQCNN-EM                   & \textbf{0.288 ± 0.021} & \textbf{92.8 ± 1.351}    & \textbf{0.298 ± 0.039} & \textbf{91.8 ± 0.671}    & \textbf{0.343 ± 0.104} & \textbf{90.1 ± 2.074}   \\
\bottomrule
\end{tabular}
\caption{Training and test classification performance of QCNN and hybrid QCNN models across circuit sizes and noise settings. Results are reported for four-, eight-, and ten-qubit circuits under noiseless and noisy simulations. Metrics include Binary Cross-Entropy (BCE) loss and classification accuracy, reported as mean $\pm$ one standard deviation over five independent training runs. Arrows indicate the preferred direction of each metric. Bold values denote the best-performing model for each device and circuit size.}
\label{tab:clf_performance}
\end{table}

Table~\ref{tab:clf_performance} presents classification performance results for all model variants across circuit sizes and noise settings. While the table reports data for four-, eight-, and ten-qubit circuits, we focus our immediate analysis on the eight-qubit configuration to complement the training dynamics observed in Figure~\ref{fig:clf_loss_convergence}. The corresponding loss trajectories for the four- and ten-qubit models are provided in Supplementary Information, confirming that the performance advantage of the hybrid models becomes more pronounced as the circuit size increases. Each metric is reported as the mean value over five independent training trials, accounting for randomness in parameter initialization, backend calibration data, and quantum simulation, accompanied by one standard deviation. The best-performing model within each device and qubit configuration is highlighted in bold.

Under the noiseless condition, all models achieve comparable accuracy. However, the proposed hybrid models---HQCNN-EZ and HQCNN-EM---yield lower BCE losses than the baseline QCNN, indicating better optimization. In particular, HQCNN-EM exhibits smaller standard deviations compared to HQCNN-EZ, reflecting more consistent training behavior. Under noisy conditions---specifically \texttt{FakeGuadalupeV2} and \texttt{AerSimulator} with \texttt{IBM\_Yonsei} backend---the baseline QCNN’s performance degrades substantially, with accuracy dropping by approximately $15$ to $20$ percentage points on average. In contrast, both hybrid models maintain high accuracy and low BCE losses, with consistently smaller standard deviations across all noise settings; notably, HQCNN-EM maintains a test accuracy of around $90\%$ under these noisy settings. Overall, the hybrid models demonstrate clear advantages in robustness to noise. Among them, HQCNN-EM shows the strongest resilience, preserving high accuracy close to the noiseless case while maintaining tighter performance variability under noise.

\subsection{Performance advantage scaling with circuit size}
\label{subsec:Performance_Advantage_w_circuit_size}
Here, we investigate the impact of increasing the quantum circuit size on model performance, specifically characterizing the comparative noise resilience of the hybrid architecture against the baseline. While larger qubit systems theoretically enable more expressive models, they inevitably incur greater noise accumulation from deeper circuits, which typically causes the performance of standard hierarchical architectures to deteriorate. We demonstrate that the proposed depth-stratified approach does not simply withstand this scaling challenge but becomes increasingly critical for performance as the system complexity grows. Specifically, our results indicate that the relative benefit of the hybrid architecture expands significantly as the circuit scale increases, effectively compensating for the growing noise volume that accompanies larger quantum systems.

To empirically validate this scaling advantage, we revisit the comprehensive results in Table~\ref{tab:clf_performance}. Under noiseless conditions, the performance of all models remains generally consistent, though the baseline QCNN exhibits a noticeable drop in accuracy for the ten-qubit configuration. However, under noisy conditions---specifically \texttt{FakeGuadalupeV2} and \texttt{AerSimulator} (\texttt{IBM\_Yonsei})---the baseline QCNN’s accuracy declines sharply as the number of qubits increases, falling by approximately 15 to 20 percentage points on average, while its BCE loss remains elevated above 0.60. In contrast, the proposed hybrid models---HQCNN-EZ and HQCNN-EM---maintain consistently high accuracy (> 85\%) and low BCE loss (< 0.50) even as the circuit size scales up to ten qubits. Notably, HQCNN-EM exhibits smaller standard deviations of performance metrics than HQCNN-EZ, indicating enhanced training stability. These findings suggest that incorporating depth-stratified mid-circuit measurements not only improves noise resilience but also ensures that the model's predictive power scales robustly with increasing circuit complexity.

\begin{figure}[ht]
\centering
\includegraphics[width=0.95\textwidth]{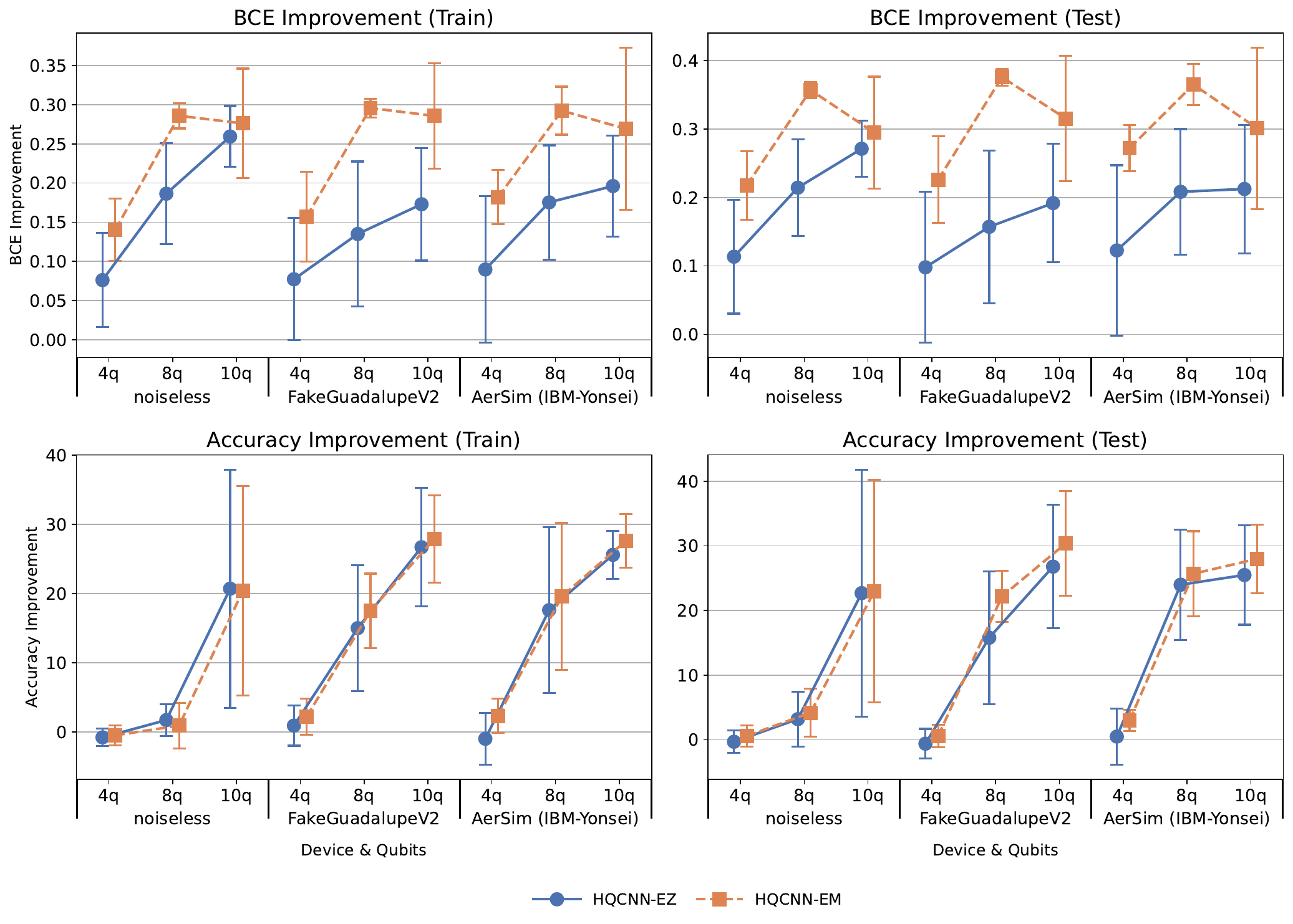}
\caption{Mean performance improvements of hybrid QCNN models relative to the baseline QCNN across circuit sizes and noise settings. Results are reported for the classification task using four-, eight-, and ten-qubit circuits (denoted as \texttt{4q}, \texttt{8q}, and \texttt{10q}) under varying simulation conditions: noiseless, \texttt{FakeGuadalupeV2}, and \texttt{AerSimulator} configured with the \texttt{IBM\_Yonsei} backend. Metrics include the relative improvement in Binary Cross-Entropy (BCE) loss (top row) and classification accuracy (bottom row), reported as mean $\pm$ one standard deviation over five independent training trials. Blue circles with solid lines denote HQCNN-EZ, while orange squares with dashed lines represent HQCNN-EM. Positive values of the relative improvement indicate a performance gain (i.e., lower loss or higher accuracy) over the baseline.}
\label{fig:clf_perf_improve}
\end{figure}

To further quantify this scaling advantage, Figure~\ref{fig:clf_perf_improve} visualizes the mean performance improvement of the hybrid models over the baseline QCNN. The relative improvement $\Delta M_k$ for the model $k$ and metric $M$ is defined as:
\begin{equation}\label{eq:improv_of_performance}
\Delta M_k = s_M \frac{1}{R}\sum_{i=1}^{R}{\bigl( M_k^{(i)} - M_{\mathrm{baseline}}^{(i)}\bigr),}
\end{equation}
where $R$ denotes the number of repeated training runs, $M_k^{(i)}$ and $M_{\mathrm{baseline}}^{(i)}$ represent the metric values for the hybrid and baseline models in the $i$-th run, respectively, and $s_M \in \{-1, 1\}$ is a sign coefficient ensuring that positive values always indicate improvements (i.e., $s_M = +1$ for accuracy and $s_M = -1$ for BCE loss).

As shown in Figure~\ref{fig:clf_perf_improve}, the improvement margin for both HQCNN-EZ and HQCNN-EM increases along with the circuit size. For BCE improvement, we observe a consistent upward trend across all noise settings: as the qubit count increases, both hybrid models yield progressively larger reductions in loss compared to the baseline. The trend for accuracy improvement varies depending on the noise setting. Under noiseless conditions, the accuracy gains for four- and eight-qubit (\texttt{q4} and \texttt{q8}) circuits are marginal, as the baseline QCNN already maintains high accuracy. However, under realistic noise, the benefits become substantial. For the four- and eight-qubit noisy regimes, the mean accuracy improvement already exceeds 15 percentage points on average. This gap amplifies significantly in the ten-qubit case (\texttt{q10}), where the baseline performance collapses; here, the hybrid models achieve their maximum relative gain, with HQCNN-EM consistently yielding the largest performance improvements. This widening performance gap confirms that the proposed architecture effectively mitigates the noise-induced degradation that hinders the performance of standard QCNNs. This suggests that the depth-stratified approach remains advantageous as the system size grows, indicating that its utility extends to regimes beyond the reach of classical simulation.

\subsection{SHAP analysis of depth-stratified measurements}
\label{subsec:SHAP_depth-stratified_meas}
The widening performance gap observed in the previous section suggests that the hybrid model effectively leverages shallow-layer measurements to stabilize performance against cumulative noise. To validate this hypothesis and confirm that performance gains stem explicitly from utilizing depth-stratified measurements, rather than simply from the additional parameters introduced by the classical neural network (architecture detailed in Section~\ref{subsec:Model_architecture}), we employ SHAP (SHapley Additive exPlanations) analysis \cite{lundberg2017unified}. Using the DeepLIFT estimator \cite{shrikumar2017learning}, we quantify the specific contribution of individual depth-stratified measurements to the final prediction, thereby isolating the impact of extracting features from shallower circuit depths.

\begin{figure}[!t]
\centering
\includegraphics[width=0.95\textwidth]{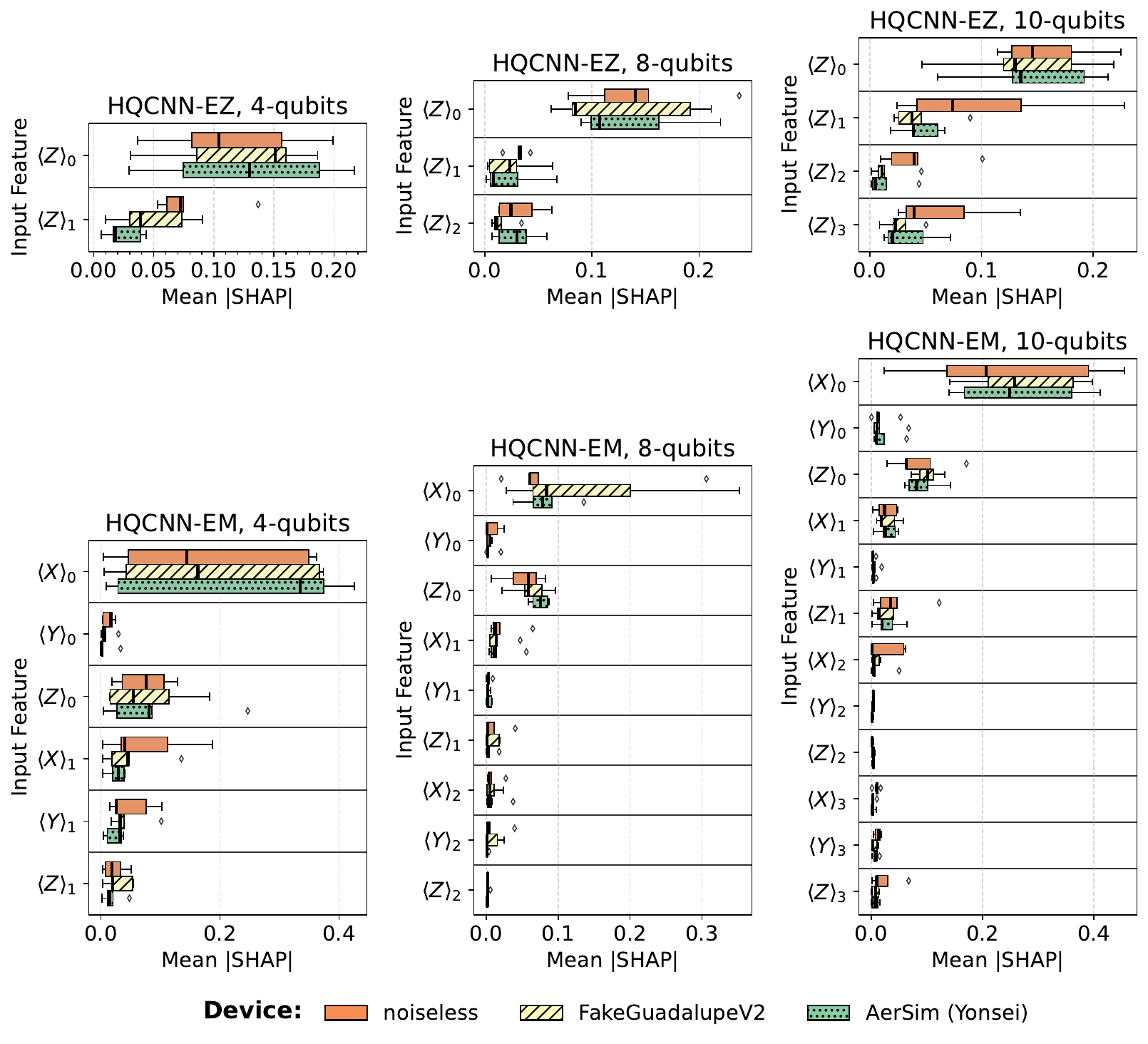}
\caption{Distribution of mean absolute SHAP values for depth-stratified measurements in the classification task across hybrid models, circuit sizes, and noise conditions. Global feature importance is quantified by the mean absolute SHAP value on the test dataset, where a higher magnitude indicates a stronger influence of the feature on the model's output. The box plots summarize the distribution of these scores across five independent trials, capturing stochastic variability. The central vertical line represents the median, serving as a robust indicator of importance. (Boxes: interquartile range; Whiskers: $1.5\times$ IQR; Points: outliers). Colors distinguish noise settings: noiseless (orange), \texttt{FakeGuadalupeV2} (yellow-hatched), and \texttt{AerSimulator} with \texttt{IBM\_Yonsei} backend (green-dotted). The features (depth-stratified measurements) on the y-axis are ordered by circuit depth from shallow (top) to deep (bottom).}
\label{fig:clf_abs_shap_boxplot}
\end{figure}

By definition, SHAP values are computed for each individual data point, assigning an importance score to every measurement outcome based on its contribution to that specific prediction. In the context of our binary classification task, a positive SHAP value indicates a contribution toward a specific label (digit 1 from MNIST), whereas a negative SHAP value indicates a contribution toward the other label (digit 0 from MNIST). However, to assess the overall importance of each qubit regardless of directionality, we first convert the raw SHAP values to their absolute magnitudes and then calculate the mean of these absolute values for each feature across the dataset. By this metric, a higher mean absolute value reflects that the specific measurement outcome exerts a stronger influence on the model's output, whereas a value near zero indicates a negligible contribution. Since parameterized models can converge to distinct weight configurations depending on initialization---leading to variations in the assigned SHAP values for identical inputs---we analyze the distribution of these mean absolute SHAP values across five independent training trials. Figure~\ref{fig:clf_abs_shap_boxplot} visualizes this distribution using box plots, where the central vertical line indicates the median importance and the box captures the interquartile range. In this figure, the features---corresponding to the depth-stratified measurements---are represented on the y-axis as the expectation value $\langle P \rangle_i$, where $P \in \{X, Y, Z\}$ denotes the measurement basis and $i$ indicates the index of the measured trash qubit, ordered by circuit depth from shallow (top) to deep (bottom).

The analysis reveals a distinct contribution pattern confirming that the proposed architecture effectively utilizes measurements from shallower circuit depths. Across all configurations, the earliest trash qubit measurements—typically those from the first pooling layer, such as $\langle Z \rangle_0$ or $\langle X \rangle_0$—consistently exhibit the highest median contribution scores. Despite the stochastic variability observed across trials (indicated by the interquartile ranges), the distribution of importance for these shallow features is systematically shifted toward higher values compared to those of deeper features (e.g., $\langle Z \rangle_2$ or $\langle Z \rangle_3$). This provides empirical evidence that qubits measured at shallower circuit depths retain more information for classification, which is often obscured by cumulative noise in standard deep hierarchical circuits. Furthermore, this dominance of early features is consistently observed across all simulation environments, from the ideal noiseless baseline to the realistic noise profiles of \texttt{FakeGuadalupeV2} and \texttt{AerSimulator} (\texttt{IBM\_Yonsei}), demonstrating that the hybrid architecture successfully identifies and leverages low-noise information pathways under diverse hardware error profiles. Notably, in the multi-basis HQCNN-EM variants, the $\langle X \rangle$ and $\langle Z \rangle$ bases show particularly strong influence compared to the $\langle Y \rangle$ basis, suggesting that the relevant features for this dataset are predominantly encoded in the $X$ and $Z$ measurement bases.

Moreover, the contribution profile demonstrates that the hybrid model utilizes a complementary combination of shallow and deep features. While the median influence of intermediate trash qubits generally decreases with circuit depth, the final remaining qubit in specific configurations exhibits a resurgence in SHAP importance, occasionally surpassing the contribution of measurements from the intermediate pooling layers (e.g., as observed in the ten-qubit HQCNN-EZ distribution). This indicates that the hybrid architecture successfully propagates and compresses relevant features to the end of the circuit, preserving the hierarchical inductive bias of the QCNN while simultaneously extracting early information. Ultimately, the distinct high importance assigned to the shallow trash qubits provides empirical evidence that the classical network is actively exploiting the depth-stratified measurement information. This confirms that the observed performance gains are not merely artifacts of increased parameter count but are driven by the active recovery of specific high-fidelity quantum features from the trash qubits.

\section{Discussion}\label{sec:Discussion}
In this work, we proposed and analyzed a noise-adaptive hybrid architecture that augments the QCNN with depth-stratified feature extraction. Hierarchical quantum classifiers represent a promising direction for capturing complex correlations using logarithmic circuit depth. However, their practical efficacy on NISQ devices and early fault-tolerant processors is often hindered by the accumulation of gate errors and decoherence along the circuit depth. By systematically recovering information from discarded trash qubits and processing these outcomes via a classical neural network, our approach effectively mitigates performance degradation caused by noise accumulation, demonstrating robust classification capabilities under realistic noise profiles derived from actual quantum processors. Furthermore, this strategy ensures compact classical post-processing; since the number of depth-stratified measurements scales logarithmically with the system size ($\mathcal{O}(\log{N})$), the classical post-processing operates on an exponentially smaller feature space rather than the full Hilbert space.

Simulation results under realistic IBM Quantum noise models demonstrate that the proposed hybrid model yields faster, more stable loss convergence and higher classification accuracy with reduced standard error compared to the standard QCNN baseline. Moreover, the analysis of classification accuracy across the number of qubits in the hierarchical quantum circuit reveals that while the baseline QCNN suffers from severe performance degradation as the number of qubits increases, the proposed hybrid variants maintain robust performance. The widening performance gap relative to the baseline underscores that leveraging measurements from shallower layers becomes critical for maintaining consistent classification accuracy under simulated noise as the hierarchical quantum circuit size grows. Indeed, our SHAP analysis validates this strategy, suggesting that the model prioritizes reliable measurements from shallow layers to mitigate the impact of cumulative noise. Ultimately, this suggests that in the presence of noise, the optimal strategy is not to rely solely on the final measurement of the hierarchical quantum circuit but to utilize less noise-affected information by measuring layer-wise trash qubits and combine this with the final measurement to yield a more accurate model prediction.

While this study primarily focuses on the classification task, the hybrid architecture is inherently adaptable to continuous real-valued targets. To validate this versatility, we applied the same depth-stratified measurement strategy to a supervised multiple regression task. As detailed in the Supplementary Information, these experiments yield results consistent with the classification task, confirming that the noise robustness of the hybrid architecture extends to regression domains.

Beyond its noise robustness, the proposed framework offers a distinct structural advantage for quantum-assisted dimensionality reduction and feature extraction. The quantum circuit component effectively functions as a trainable encoder that compresses the high-dimensional state into a compact classical vector composed of the depth-stratified measurement outcomes. This allows the downstream classical neural network to operate with significantly reduced complexity; specifically, the input dimension of the classical head scales logarithmically with the number of qubits. For demonstration purposes, this study employs angle encoding, which maps $N$ classical features onto $N$ qubits. However, the proposed architecture is agnostic to the specific encoding scheme and can readily accommodate alternative encodings. For example, amplitude encoding could be used to further enhance the compression efficiency of the architecture. Since amplitude encoding maps $N$ features to $\bigl\lceil \log_2 (N) \bigr\rceil$ qubits, integrating it with a QCNN employing the depth-stratified measurement scheme could theoretically achieve a double-logarithmic reduction $(\mathcal{O}({\log({\log{N}}})))$ in the classical model size relative to the raw input dimension. This highlights the potential of the QCNN architecture not only as a standalone classifier but also as a highly compact feature extractor for high-dimensional classical data.

Several avenues for future research remain to optimize the proposed depth-stratified framework. First, regarding the measurements strategy, our current approach selects measurement locations heuristically---specifically, measuring the first-indexed qubit at each pooling layer. While this approach was sufficient to demonstrate significant noise robustness and performance improvements, future work could investigate adaptive measurement strategies guided by circuit-specific analysis to identify the most informative qubits. This strategy can be extended to optimizing the number of trash qubits being measured; however, one must consider that increasing the number of measurements may increase the capacity of the classical part of the hybrid architecture. If the classical component becomes dominant, performance gains might arise from classical correlations rather than quantum processing, potentially eliminating the quantum advantage. Second, future research should explore advanced architectures for both the classical and quantum components. On the classical side, integrating sequential models like Recurrent Neural Networks (RNNs) could better exploit layer-wise correlations compared to simple feed-forward neural networks. On the quantum side, incorporating hierarchical quantum models, such as Tree Tensor Networks (TTN) or the Multi-scale Entanglement Renormalization Ansatz (MERA), would broaden the applicability of this hybrid architecture.

In conclusion, this study establishes depth-stratified feature extraction as a viable design principle for noise-robust quantum machine learning. By reinterpreting discarded qubits as informative resources, it offers a practical pathway for improving the reliability of hierarchical quantum models on noisy quantum processors.

\section{Methods}\label{sec:Methods}

\subsection{Data Description and Encoding Strategy}
\label{subsec:data_description_and_encoding_strategy}
For the classification task, we used the MNIST dataset, consisting of $28 \times 28$ grayscale images of handwritten digits. We formulated a binary classification problem distinguishing between digits $0$ and $1$. The data is partitioned into strictly balanced subsets of 200 training, 50 validation, and 200 test samples. Formally, we train a parameterized quantum model on data $\{(x_i, y_i)\}_{i=1}^{N}$ where $y_i \in \{0, 1\}$ to predict the class label from the encoded quantum state.

To encode classical data into an $n$-qubit circuit, we employ angle encoding, also known as qubit encoding, where an $n$-dimensional feature vector $\mathbf{x}\in\mathbb{R}^n$ is mapped to the rotation angles of single-qubit $RY$ gates. Since the raw input dimensionality ($28 \times 28 = 784$) exceeds the available qubit count, we apply Principal Component Analysis (PCA) to reduce the image data into an $n$-dimensional feature vector. These extracted features are then rescaled to the interval $[0, \pi]$ to ensure they map effectively to the qubit rotation space. For the varying circuit configurations of $n\in\{4,8,10\}$ qubits, we retain the top $n$ principal components and encode them via these parameterized rotations.

We deliberately chose this minimal encoding scheme over more complex entangling maps (e.g., the ZZ feature map~\cite{havlivcek2019supervised}) to minimize state preparation overhead. Specifically, whereas angle encoding relies exclusively on single-qubit $RY$ rotation gates, the ZZ feature map necessitates the use of two-qubit CNOT gates, which inevitably accumulates two-qubit errors---a dominant source of noise in NISQ devices---before the variational quantum circuit even begins. This additional noise at the encoding stage could severely degrade the input state such that random noise dominates, thereby compromising the effectiveness of the subsequent variational circuit. By adhering to a depth-efficient encoding strategy, we ensure that any observed improvements in noise robustness are attributed to the proposed depth-stratified feature extraction rather than the properties of the data embedding.

\subsection{Hardware and Software Setup}\label{subsec:Hardware_software_setup}
All computational experiments were conducted on the Windows 11 operating system using two hardware configurations: (1) an Intel Core i9-12900H CPU (14 cores) with 32 GB RAM, and (2) an Intel Core i9-12900KF CPU (16 cores) with 64 GB RAM. The software framework was developed in Python. Quantum circuit simulations were executed using IBM Qiskit, utilizing the Aer provider for noise modeling, while the classical neural network architecture and optimization routines were implemented in PyTorch.

\subsection{Noisy Simulation Settings}
\label{subsec:Quantum_noisy_simul_env}
To evaluate model performance under different noise conditions, we employed three different quantum backends, each representing a distinct noise profile. When targeting a specific IBM Quantum backend, we transpile the circuit using Qiskit’s preset transpilation pipeline with optimization level 1, which applies light optimization—including qubit layout selection and routing to satisfy the device coupling constraints. Specifically, the transpiler decomposes the logical circuit into the backend's native basis gate set: $\{RZ, \sqrt{X}, X, CX\}$ for the snapshot-based backend and $\{RZ, \sqrt{X}, X, ECR\}$ for the \texttt{IBM\_Yonsei} device. Subsequently, noise processes---including gate and measurement errors---are incorporated according to the backend calibration data.

For the classification task, we executed circuits with 256 shots per measurement. This value was selected to balance the precision of expectation value estimation with the computational feasibility of the training process, which requires extensive circuit executions for gradient calculation via the parameter-shift rule.

The first backend, the noiseless simulator, provides an idealized quantum execution setting without any noise sources and serves as a reference for the ideal model performance achievable in the absence of hardware noise. The second backend, the snapshot-based simulator, mimics the behavior of previously available IBM Quantum systems using stored system snapshots that include hardware characteristics such as the coupling map and qubit properties (e.g., $T_1$, $T_2$, and gate error rates). For this category, we utilized the \texttt{FakeGuadalupeV2} backend. The third backend, the \texttt{AerSimulator} with a real-device noise model, reproduces the noise characteristics of an existing IBM Quantum device by importing its live calibration data. In this work, we loaded the system noise properties from the \texttt{IBM\_Yonsei} device---IBM’s Eagle r3 quantum processor. 

For each simulation, we have recorded the hardware calibration metrics of the physical qubits assigned to the transpiled circuits for all three models (QCNN, HQCNN-EZ, and HQCNN-EM). The \textit{Calibration Data} section in the Supplementary Information provides the mean and the standard deviation of all the backend properties utilized in the simulation, including coherence times ($T_1$, $T_2$), qubit frequencies, anharmonicity, and error profiles for single-qubit gates (e.g., $SX$, $X$, $RZ$), two-qubit gates ($CNOT, ECR$), and readout operations.

\subsection{Model Architecture}
\label{subsec:Model_architecture}

\begin{figure}[htbp]
    \centering
    \includegraphics[width=1.0\textwidth]{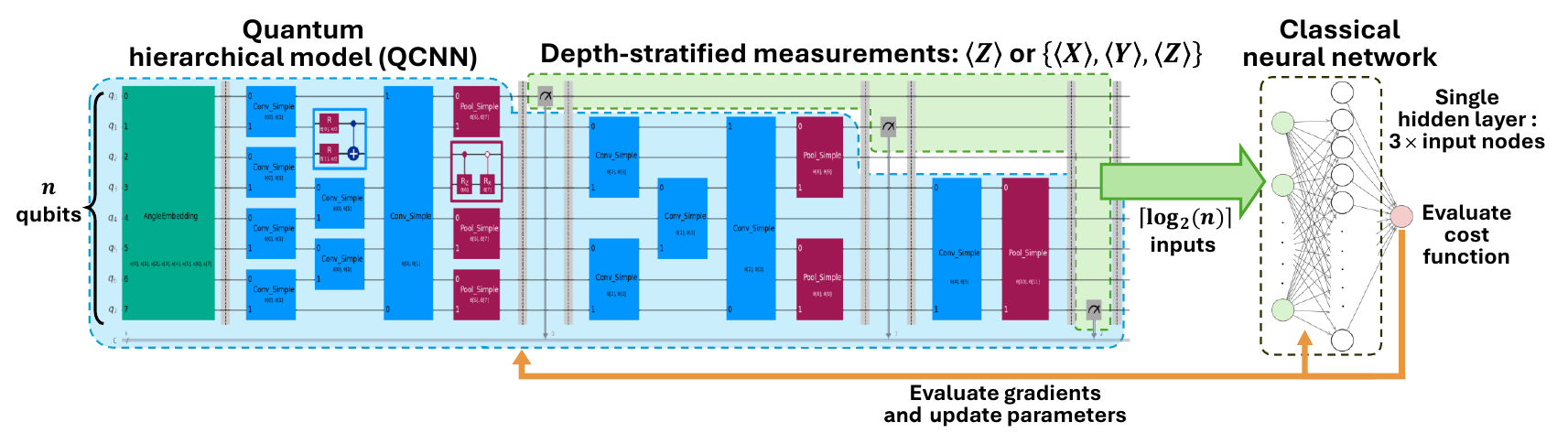}
    \caption{Schematic of the proposed eight-qubit hybrid QCNN architecture, illustrated using an eight-qubit circuit. The framework integrates a hierarchical quantum circuit (left blue block) with a classical neural network (right dashed block) through depth-stratified measurements (green block). Qubits discarded during the pooling layers are measured to extract expectation values---$\langle Z \rangle$ for HQCNN-EZ, or $\bigl\{\langle X \rangle, \langle Y \rangle, \langle Z \rangle \bigr\}$ for HQCNN-EM---which, together with the final qubit readout, form the input vector to the classical component. The classical network processes these inputs through a single hidden layer architecture to compute the final prediction.}
    \label{fig:hqcnn_circuit}
\end{figure}

There exist multiple ways to design the convolutional and pooling filters in a QCNN circuit. In this work, as illustrated in Figure \ref{fig:hqcnn_circuit}, we adopt a simple yet representative configuration: the convolutional filter consists of two single-qubit $RY$ rotation gates followed by one $CNOT$ gate, while the pooling filter comprises one controlled-$RZ$ and one controlled-$RX$ rotation gate. This minimal architecture was deliberately chosen to isolate and highlight the effect of depth-stratified measurements on the trash qubits---measured either in the computational ($Z$) basis or in multiple bases ($X$, $Y$, and $Z$)---without introducing unnecessary architectural complexity.

We evaluated three distinct model configurations. The first is the standard QCNN, which serves as the baseline by generating predictions solely from the final qubit measurement. The remaining two models are hybrid variants (HQCNNs) designed to recover information via depth-stratified feature extraction. The first variant, HQCNN-EZ, measures the expectation value $\langle Z \rangle$ of the discarded qubits. The second variant, HQCNN-EM, extends this protocol to a multi-basis scheme, measuring expectation values $\bigl\{ \langle X \rangle, \langle Y \rangle, \langle Z \rangle \bigr\}$ of each trash qubit, thereby capturing a richer representation of the quantum state and tripling the dimensionality of the classical input vector compared to the EZ variant.

The classical neural network component in the hybrid models is intentionally restricted to an architecture with a single hidden layer. Let $f_\theta : \mathbb{R}^{m(n)} \to [0,1]$ denote the classical neural network with parameters $\theta$, defined as
\begin{equation*}
	f_\theta(\mathbf{x})=
	\sigma \left(
		\mathbf{W}_2 \; \mathrm{ReLU} \left(
			\mathbf{W}_1 \mathbf{x} + \mathbf{b}_1
		\right) +
		\mathbf{b}_2
	\right),
\end{equation*}
where $\sigma$ denotes the sigmoid activation function, $m(n)$ is the total number of measurement outcomes, and
\begin{equation*}
    \mathbf{W}_1 \in \mathbb{R}^{3m(n) \times m(n)}, \quad
    \mathbf{W}_2 \in \mathbb{R}^{1 \times 3m(n)}.
\end{equation*}

This minimal design limits the capacity of the classical neural network, preventing it from overfitting or dominating the learning process---ensuring the task is not solved solely by the classical component---while effectively utilizing the information from the depth-stratified measurements to enhance the model's robustness against noise. The number of input nodes corresponds to the total number of measurement outcomes, which aggregates both the depth-stratified trash measurements and the final qubit readout. Consequently, the structural complexity of the classical network adapts to the measurement scheme: for HQCNN-EZ, the number of input nodes is equal to $\bigl\lceil \log_2(n)\bigr\rceil$, where $n$ is the number of qubits, whereas for HQCNN-EM, the number of input nodes is three times that of HQCNN-EZ. The hidden layer contains three times as many nodes as the input layer and utilizes the ReLU activation function. The output layer consists of a single node with a sigmoid activation function to produce the final binary classification probability. Importantly, this hybrid architecture is constructed as a unified differentiable pipeline, meaning the quantum circuit parameters and classical neural network weights are trained jointly in an end-to-end manner.

\subsection{Performance Metrics}
\label{subsec:Performance_Metrics}
To evaluate the classification performance, we utilize two standard metrics: Binary Cross-Entropy (BCE) loss and classification accuracy. The BCE loss measures the divergence between the predicted probability $\hat{y}_i$ and the true binary label $y_i \in \{0, 1\}$. It is defined as
\begin{equation}
    \text{BCE loss} = -\frac{1}{M} \sum_{i=1}^M \bigl( y_i \log(\hat{y}_i) + (1 - y_i) \log(1 - \hat{y}_i) \bigr),
\end{equation}
where $M$ denotes the total number of samples. A lower BCE loss indicates that the model's predicted probabilities are more closely aligned with the true class labels.

Classification accuracy quantifies the overall correctness of the model's predictions and is calculated as:
\begin{equation}
    \text{Accuracy} = \frac{1}{M} \sum_{i=1}^M \mathds{1}(\hat{y}_i = y_i),
\end{equation}
where the predicted label is obtained by thresholding the probability $\hat{y}_i$ at $0.5$, and $\mathds{1}(\cdot)$ is the indicator function that returns $1$ if the condition is satisfied and $0$ otherwise.

\bibliography{ref}

@article{biamonte2017quantum,
  title={Quantum machine learning},
  author={Biamonte, Jacob and Wittek, Peter and Pancotti, Nicola and Rebentrost, Patrick and Wiebe, Nathan and Lloyd, Seth},
  journal={Nature},
  volume={549},
  number={7671},
  pages={195--202},
  year={2017},
  publisher={Nature Publishing Group UK London}
}

@article{harrow2017quantum,
  title={Quantum computational supremacy},
  author={Harrow, Aram W and Montanaro, Ashley},
  journal={Nature},
  volume={549},
  number={7671},
  pages={203--209},
  year={2017},
  publisher={Nature Publishing Group}
}

@article{schuld2019quantum,
  title={Quantum machine learning in feature Hilbert spaces},
  author={Schuld, Maria and Killoran, Nathan},
  journal={Physical review letters},
  volume={122},
  number={4},
  pages={040504},
  year={2019},
  publisher={APS}
}

@article{rebentrost2014quantum,
  title={Quantum support vector machine for big data classification},
  author={Rebentrost, Patrick and Mohseni, Masoud and Lloyd, Seth},
  journal={Physical review letters},
  volume={113},
  number={13},
  pages={130503},
  year={2014},
  publisher={APS}
}

@article{orus2014practical,
  title={A practical introduction to tensor networks: Matrix product states and projected entangled pair states},
  author={Or{\'u}s, Rom{\'a}n},
  journal={Annals of physics},
  volume={349},
  pages={117--158},
  year={2014},
  publisher={Elsevier}
}

@article{shi2006classical,
  title = {Classical simulation of quantum many-body systems with a tree tensor network},
  author = {Shi, Y.-Y. and Duan, L.-M. and Vidal, G.},
  journal = {Phys. Rev. A},
  volume = {74},
  issue = {2},
  pages = {022320},
  numpages = {4},
  year = {2006},
  month = {Aug},
  publisher = {American Physical Society},
  doi = {10.1103/PhysRevA.74.022320},
  url = {https://link.aps.org/doi/10.1103/PhysRevA.74.022320}
}

@article{verstraete2008matrix,
  title={Matrix product states, projected entangled pair states, and variational renormalization group methods for quantum spin systems},
  author={Verstraete, Frank and Murg, Valentin and Cirac, J Ignacio},
  journal={Advances in physics},
  volume={57},
  number={2},
  pages={143--224},
  year={2008},
  publisher={Taylor \& Francis}
}

@article{grant2018hierarchical,
  title={Hierarchical quantum classifiers},
  author={Grant, Edward and Benedetti, Marcello and Cao, Shuxiang and Hallam, Andrew and Lockhart, Joshua and Stojevic, Vid and Green, Andrew G and Severini, Simone},
  journal={npj Quantum Information},
  volume={4},
  number={1},
  pages={65},
  year={2018},
  publisher={Nature Publishing Group UK London}
}

@article{vidal2008class,
  title = {Class of Quantum Many-Body States That Can Be Efficiently Simulated},
  author = {Vidal, G.},
  journal = {Phys. Rev. Lett.},
  volume = {101},
  issue = {11},
  pages = {110501},
  numpages = {4},
  year = {2008},
  month = {Sep},
  publisher = {American Physical Society},
  doi = {10.1103/PhysRevLett.101.110501},
  url = {https://link.aps.org/doi/10.1103/PhysRevLett.101.110501}
}

@article{cong2019quantum,
  title={Quantum convolutional neural networks},
  author={Cong, Iris and Choi, Soonwon and Lukin, Mikhail D},
  journal={Nature Physics},
  volume={15},
  number={12},
  pages={1273--1278},
  year={2019},
  publisher={Nature Publishing Group UK London}
}

@article{pesah2021absence,
  title={Absence of barren plateaus in quantum convolutional neural networks},
  author={Pesah, Arthur and Cerezo, Marco and Wang, Samson and Volkoff, Tyler and Sornborger, Andrew T and Coles, Patrick J},
  journal={Physical Review X},
  volume={11},
  number={4},
  pages={041011},
  year={2021},
  publisher={APS}
}

@article{hur2022quantum,
  title={Quantum convolutional neural network for classical data classification},
  author={Hur, Tak and Kim, Leeseok and Park, Daniel K},
  journal={Quantum Machine Intelligence},
  volume={4},
  number={1},
  pages={3},
  year={2022},
  publisher={Springer}
}

@article{preskill2018quantum,
  title={Quantum computing in the NISQ era and beyond},
  author={Preskill, John},
  journal={Quantum},
  volume={2},
  pages={79},
  year={2018},
  publisher={Verein zur F{\"o}rderung des Open Access Publizierens in den Quantenwissenschaften}
}

@article{lee2023scalable,
  title={Scalable quantum measurement error mitigation via conditional independence and transfer learning},
  author={Lee, Changwon and Park, Daniel K},
  journal={Machine Learning: Science and Technology},
  volume={4},
  number={4},
  pages={045051},
  year={2023},
  publisher={IOP Publishing}
}

@article{zapletal2024error,
  title={Error-tolerant quantum convolutional neural networks for symmetry-protected topological phases},
  author={Zapletal, Petr and McMahon, Nathan A and Hartmann, Michael J},
  journal={Physical Review Research},
  volume={6},
  number={3},
  pages={033111},
  year={2024},
  publisher={APS}
}

@article{ahmed2025quantum,
  title={Quantum neural networks: A comparative analysis and noise robustness evaluation},
  author={Ahmed, Tasnim and Kashif, Muhammad and Marchisio, Alberto and Shafique, Muhammad},
  journal={arXiv preprint arXiv:2501.14412},
  year={2025}
}

@article{deng2012mnist,
  title={The mnist database of handwritten digit images for machine learning research},
  author={Deng, Li},
  journal={IEEE Signal Processing Magazine},
  volume={29},
  number={6},
  pages={141--142},
  year={2012},
  publisher={IEEE}
}

@article{lundberg2017unified,
  title={A unified approach to interpreting model predictions},
  author={Lundberg, Scott M and Lee, Su-In},
  journal={Advances in neural information processing systems},
  volume={30},
  year={2017}
}

@inproceedings{shrikumar2017learning,
  title={Learning important features through propagating activation differences},
  author={Shrikumar, Avanti and Greenside, Peyton and Kundaje, Anshul},
  booktitle={International conference on machine learning},
  pages={3145--3153},
  year={2017},
  organization={PMlR}
}

@article{schuld2020circuit,
  title={Circuit-centric quantum classifiers},
  author={Schuld, Maria and Bocharov, Alex and Svore, Krysta M and Wiebe, Nathan},
  journal={Physical Review A},
  volume={101},
  number={3},
  pages={032308},
  year={2020},
  publisher={APS}
}

@article{cerezo2021variational,
  title={Variational quantum algorithms},
  author={Cerezo, Marco and Arrasmith, Andrew and Babbush, Ryan and Benjamin, Simon C and Endo, Suguru and Fujii, Keisuke and McClean, Jarrod R and Mitarai, Kosuke and Yuan, Xiao and Cincio, Lukasz and others},
  journal={Nature Reviews Physics},
  volume={3},
  number={9},
  pages={625--644},
  year={2021},
  publisher={Nature Publishing Group UK London}
}

@article{benedetti2019parameterized,
  title={Parameterized quantum circuits as machine learning models},
  author={Benedetti, Marcello and Lloyd, Erika and Sack, Stefan and Fiorentini, Mattia},
  journal={Quantum science and technology},
  volume={4},
  number={4},
  pages={043001},
  year={2019},
  publisher={IOP Publishing}
}

@article{banchi2021generalization,
  title={Generalization in quantum machine learning: A quantum information standpoint},
  author={Banchi, Leonardo and Pereira, Jason and Pirandola, Stefano},
  journal={PRX Quantum},
  volume={2},
  number={4},
  pages={040321},
  year={2021},
  publisher={APS}
}

@article{jaderberg2024let,
  title={Let quantum neural networks choose their own frequencies},
  author={Jaderberg, Ben and Gentile, Antonio A and Berrada, Youssef Achari and Shishenina, Elvira and Elfving, Vincent E},
  journal={Physical Review A},
  volume={109},
  number={4},
  pages={042421},
  year={2024},
  publisher={APS}
}

@article{kim2025expressivity,
  title={Expressivity of deterministic quantum computation with one qubit},
  author={Kim, Yujin and Park, Daniel K},
  journal={Physical Review A},
  volume={111},
  number={2},
  pages={022429},
  year={2025},
  publisher={APS}
}

@article{hur2024understanding,
  title={Understanding generalization in quantum machine learning with margins},
  author={Hur, Tak and Park, Daniel K},
  journal={arXiv preprint arXiv:2411.06919},
  year={2024}
}

@article{lloyd2014quantum,
  title={Quantum principal component analysis},
  author={Lloyd, Seth and Mohseni, Masoud and Rebentrost, Patrick},
  journal={Nature physics},
  volume={10},
  number={9},
  pages={631--633},
  year={2014},
  publisher={Nature Publishing Group UK London}
}

@article{cross2015quantum,
  title={Quantum learning robust against noise},
  author={Cross, Andrew W and Smith, Graeme and Smolin, John A},
  journal={Physical Review A},
  volume={92},
  number={1},
  pages={012327},
  year={2015},
  publisher={APS}
}

@article{lourens2023hierarchical,
  title={Hierarchical quantum circuit representations for neural architecture search},
  author={Lourens, Matt and Sinayskiy, Ilya and Park, Daniel K and Blank, Carsten and Petruccione, Francesco},
  journal={npj Quantum Information},
  volume={9},
  number={1},
  pages={79},
  year={2023},
  publisher={Nature Publishing Group UK London}
}

@article{havlivcek2019supervised,
  title={Supervised learning with quantum-enhanced feature spaces},
  author={Havl{\'\i}{\v{c}}ek, Vojt{\v{e}}ch and C{\'o}rcoles, Antonio D and Temme, Kristan and Harrow, Aram W and Kandala, Abhinav and Chow, Jerry M and Gambetta, Jay M},
  journal={Nature},
  volume={567},
  number={7747},
  pages={209--212},
  year={2019},
  publisher={Nature Publishing Group UK London}
}

@article{lee2025optimizing,
  title={Optimizing quantum convolutional neural network architectures for arbitrary data dimension},
  author={Lee, Changwon and Araujo, Israel F and Kim, Dongha and Lee, Junghan and Park, Siheon and Ryu, Ju-Young and Park, Daniel K},
  journal={Frontiers in Physics},
  volume={13},
  pages={1529188},
  year={2025},
  publisher={Frontiers Media SA}
}


\section*{Acknowledgments}
This work was supported by the Yonsei University Research Grant of 2025, Institute of Information \& communications Technology Planning \& evaluation (IITP) grant funded by the Korea government (No. 2019-0-00003, Research and Development of Core Technologies for Programming, Running, Implementing and Validating of Fault-Tolerant Quantum Computing System), the National Research Foundation of Korea (RS-2025-02309510), the Ministry of Trade, Industry, and Energy (MOTIE), Korea, under the Industrial Innovation Infrastructure Development Project (RS-2024-00466693), and Korean ARPA-H Project through the Korea Health Industry Development Institute (KHIDI), funded by the Ministry of Health \& Welfare, Korea (RS-2025-25456722).

\section*{Author contributions statement}

D.K.P. conceived the study, designed the experiments, and supervised the project. T.K. and I.F.A. designed and performed the experiments. All authors contributed to the analysis, interpretation, and discussion of results, and to the writing of the manuscript.

\section*{Data availability}
The datasets and code used in this study are available at \href{https://github.com/qDNA-yonsei/Noise-Adaptive-HQCNN}
{\nolinkurl{https://github.com/qDNA-yonsei/Noise-Adaptive-HQCNN}}.

\section*{Additional information}

\textbf{Competing interests}
The authors declare no competing interests.


\section*{\Large Supplementary Information}\label{SI:Supplementary_Info}

\setcounter{figure}{0}                         
\renewcommand{\figurename}{Supplementary Figure}       
\setcounter{table}{0}                         
\renewcommand{\tablename}{Supplementary Table}       
\setcounter{section}{0}
\renewcommand{\thesection}{S\arabic{section}}

\subsection*{Noise robustness verification on supervised regression task}\label{SI:sec:Regression_task}

\subsubsection*{Synthetic data generation and task formulation}\label{SI:subsec:Data_and_task_for_reg}
To demonstrate the versatility of the proposed architecture beyond classification, we evaluate its performance on a supervised multiple linear regression task. We constructed a synthetic dataset using the \texttt{make\_regression} function from the Scikit-learn library, generating $m \in \{4, 8\}$ informative predictors sampled from independent normal distributions. The target variable $y$ was generated according to a linear model with additive Gaussian noise:
\begin{equation}
    y = \sum_{i=1}^{m}{\beta_i x_i} + \varepsilon, \quad \varepsilon \sim \mathcal{N}(0, 1),
\end{equation}
where $\beta_i$ represents the regression coefficients. A total of 220 samples were generated and partitioned into subsets of 100 training, 20 validation, and 100 test samples. To ensure compatibility with the quantum angle encoding scheme and the bounded output range of the quantum circuit, all input features were rescaled to the interval $[0, \pi]$ and the target values to the range $[-1, 1]$ after generation.

The training configuration adopted the same general architecture and protocol as the classification task to facilitate fair comparison, with specific adjustments for the regression objective. While the batch size (10) and optimizer (ADAM with a learning rate of 0.01) were kept consistent, the training duration was set to 30 epochs to ensure adequate convergence for the continuous target estimation. The classical neural network component in the hybrid models retained the same architecture used in classification: a single hidden layer with nodes equal to three times the input dimension and ReLU activation. However, the final output node utilized a tanh activation function to produce bounded real-valued predictions consistent with the normalized target range.

\subsubsection*{Training dynamics and regression performance}\label{SI:subsec:Performance_reg}
To quantify the model's predictive accuracy, we employed the Mean Squared Error (MSE) as the objective function during training. The MSE measures the average squared difference between the predicted output values of the model $\hat{y}_i$ and the true output values $y_i$, defined as
\begin{equation}
\textbf{MSE} = \frac{1}{n} \sum_{i=1}^n (y_i - \hat{y}_i)^2,
\end{equation}
where lower values indicate better predictive accuracy. Additionally, we evaluated the coefficient of determination ($R^2$) to measure the proportion of variance in the target variable explained by the model,
\begin{equation}
R^2 = 1 - \frac{\sum{i=1}^{n}(y_i - \hat{y_i})^2}{\sum_{i=1}^{n}(y_i - \bar{y})^2},
\end{equation}
where $\bar{y}$ denotes the mean of the observed target values, with values closer to $1$ indicating a stronger fit. These metrics serve as our primary benchmarks for evaluating prediction accuracy and model fit. Consistent with the classification task, we evaluated the final model performance at the point where the validation loss was lowest. We then computed the MSE and $R^2$ scores on both the training and test datasets using this best-performing version of the model.

\begin{figure}[ht]
    \centering
    \begin{subfigure}{0.9\textwidth}  
        \centering
        \includegraphics[width=\linewidth]{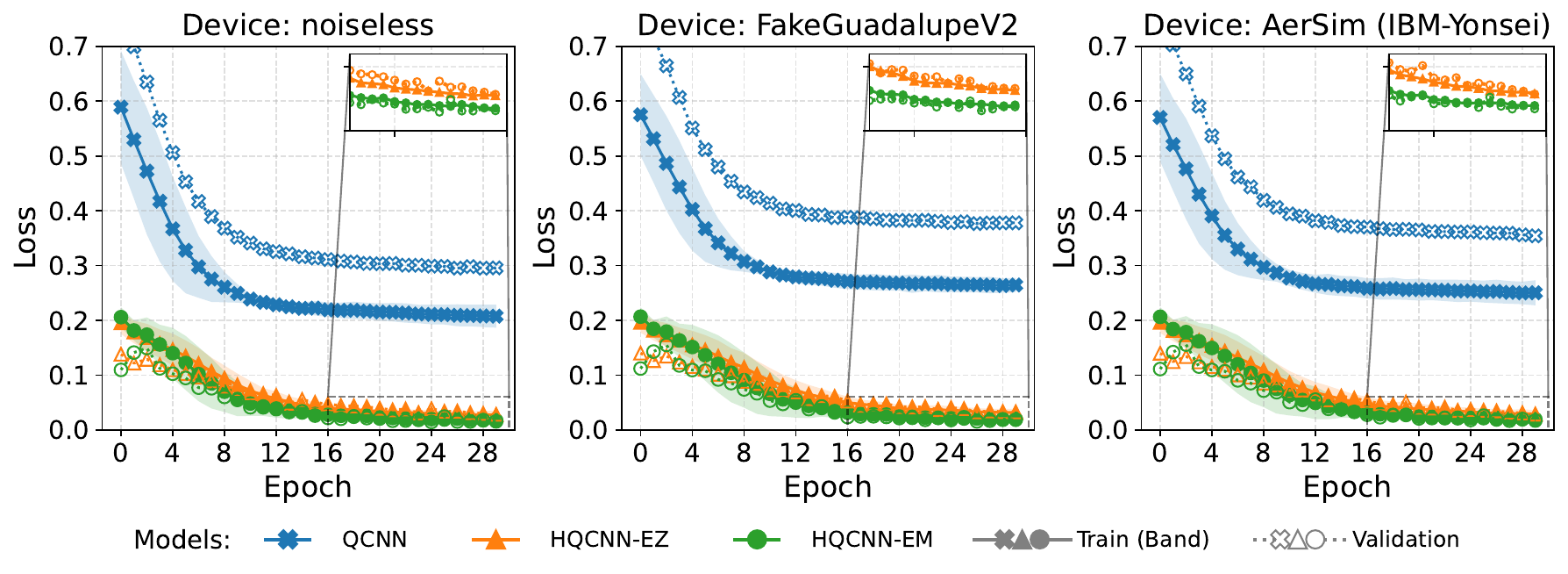}
        \caption{Training loss of four-qubit circuit}
        \label{SI_fig:reg_loss_convergence_q4}
    \end{subfigure}

    \vspace{0.8em}  

    \begin{subfigure}{0.9\textwidth}
        \centering
        \includegraphics[width=\linewidth]{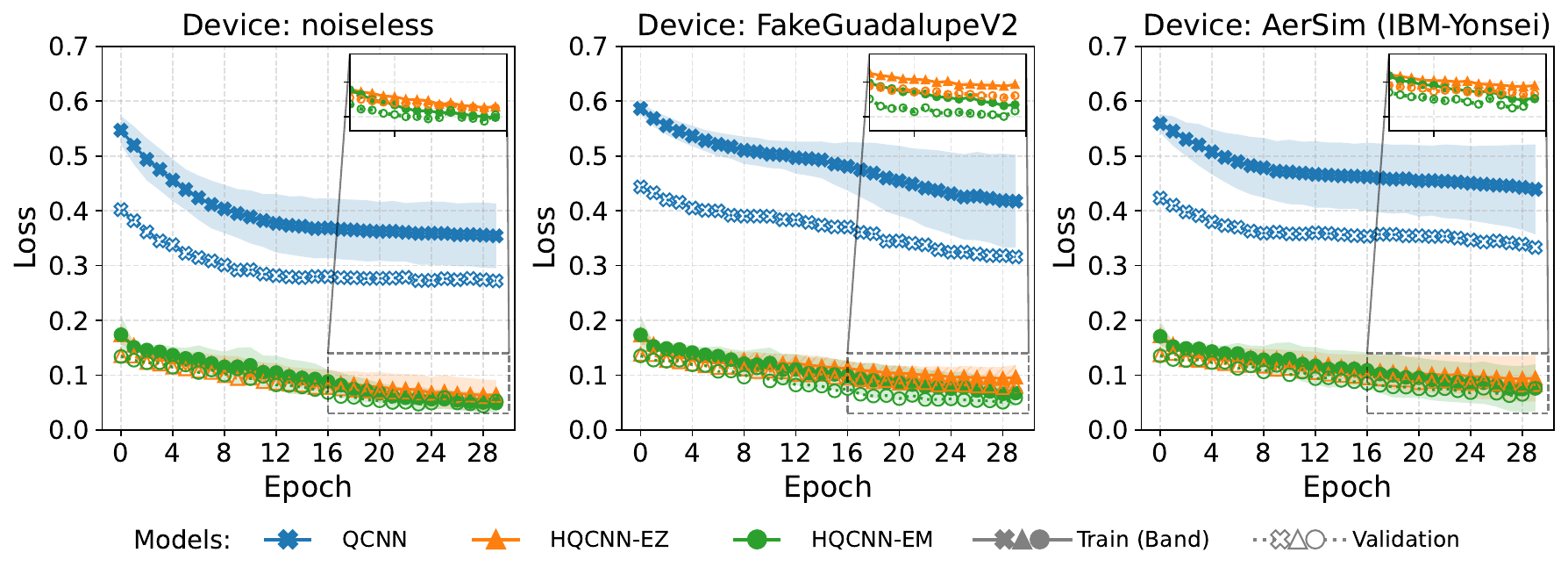}
        \caption{Training loss of eight-qubit circuit}
        \label{SI_fig:reg_loss_convergence_q8}
    \end{subfigure}
    
    \caption{Training and validation loss curves of the regression models under varying simulation conditions. The plots display the loss trajectories for (a) a four-qubit circuit and (b) an eight-qubit circuit across three environments: noiseless, \texttt{FakeGuadalupeV2}, and \texttt{AerSimulator} (configured with the \texttt{IBM\_Yonsei} backend). Solid lines with filled markers represent the mean training loss, while dotted lines with open markers indicate the mean validation loss, averaged over five independent iterations. The shaded bands around the training loss curves indicate one standard deviation ($\pm \sigma$) around the mean. The insets provide a magnified view of the final training epochs to illustrate the convergence details of the hybrid models.}
    \label{SI_fig:reg_loss_convergence}
\end{figure}

Supplementary Figure~\ref{SI_fig:reg_loss_convergence} visualizes the training and validation MSE loss trajectories for four- and eight-qubit models across three distinct simulation environments: noiseless, \texttt{FakeGuadalupeV2}, and \texttt{AerSimulator} configured with \texttt{IBM\_Yonsei} backend. The baseline QCNN exhibits limited learning capability in this regression context. While its loss decreases initially, it converges to significantly higher final values compared to the hybrid variants. This underperformance is anticipated, as the standard QCNN architecture is primarily designed for classification tasks rather than continuous value estimation; consequently, the final qubit alone fails to capture sufficient information to accurately reconstruct the continuous target variable. In contrast, both hybrid architectures (HQCNN-EZ and HQCNN-EM) demonstrate better training trajectories from the onset, starting from lower values. Even at the initial epoch, these models achieve MSE loss values substantially lower than the converged state of the baseline QCNN. Throughout the training process, they demonstrate stable convergence, consistently reaching minimal loss values across all noise settings. Notably, the performance of the HQCNN-EZ model is comparable to that of the HQCNN-EM model, although the latter converges to slightly lower final loss values in most configurations. Overall, these results demonstrate that the hybrid architectures possess enhanced trainability and robustness, maintaining high performance consistent across both four- and eight-qubit configurations.

These visual trends are quantitatively confirmed in Supplementary Table~\ref{SI_tab:reg_performance}. The baseline QCNN consistently yields significantly higher MSE values and negative $R^2$ scores across all configurations (ranging from $-0.073$ to $-3.001$), indicating that the model fails to capture the underlying regression function and performs worse than a trivial predictor that outputs the mean of the target variable. In contrast, the hybrid models attain substantially lower MSE values with reduced variance compared to the baseline. Conversely, both hybrid models successfully learn the regression task, though the performance scales differently with circuit size. For the four-qubit models, the hybrids achieve high $R^2$ scores, averaging between $0.806$ and $0.907$, alongside low MSE values. In the eight-qubit case, the models yield moderately positive $R^2$ scores averaging between $0.289$ and $0.671$. Regarding model variants, we observe that in the four-qubit regime, the HQCNN-EM model exhibits a distinct performance advantage. However, in the eight-qubit case, the HQCNN-EZ and HQCNN-EM variants display comparable performance levels, albeit with reduced predictive accuracy relative to the four-qubit configuration.

\begin{table}[ht]
\centering
\small 
\setlength{\tabcolsep}{3pt} 
\begin{tabular}{cll|cc|cc}
\toprule
\multicolumn{3}{c|}{\textbf{Regression Task}}                                                                                                                                  & \multicolumn{2}{c|}{four qubits}                & \multicolumn{2}{c}{eight qubits}                \\ \hline
\multicolumn{1}{c|}{Dataset}                & \multicolumn{1}{c|}{Device}                                                                         & \multicolumn{1}{c|}{Model} & MSE $\downarrow$       & R-squared $\uparrow$   & MSE $\downarrow$       & R-squared $\uparrow$   \\ \hline
\multicolumn{1}{c|}{\multirow{9}{*}{Train}} & \multicolumn{1}{l|}{\multirow{3}{*}{noiseless}}                                                     & QCNN                       & 0.209 ± 0.019          & -0.073 ± 0.095         & 0.361 ± 0.066          & -1.442 ± 0.445         \\
\multicolumn{1}{c|}{}                       & \multicolumn{1}{l|}{}                                                                               & HQCNN-EZ                   & 0.027 ± 0.004          & 0.861 ± 0.023          & 0.064 ± 0.025          & 0.564 ± 0.167          \\
\multicolumn{1}{c|}{}                       & \multicolumn{1}{l|}{}                                                                               & HQCNN-EM                   & \textbf{0.018 ± 0.004} & \textbf{0.907 ± 0.019} & \textbf{0.049 ± 0.019} & \textbf{0.671 ± 0.131} \\ \cline{2-7} 
\multicolumn{1}{c|}{}                       & \multicolumn{1}{l|}{\multirow{3}{*}{FakeGuadalupeV2}}                                               & QCNN                       & 0.264 ± 0.012          & -0.355 ± 0.064         & 0.414 ± 0.087          & -1.802 ± 0.587         \\
\multicolumn{1}{c|}{}                       & \multicolumn{1}{l|}{}                                                                               & HQCNN-EZ                   & 0.030 ± 0.006          & 0.847 ± 0.032          & 0.093 ± 0.021          & 0.368 ± 0.142          \\
\multicolumn{1}{c|}{}                       & \multicolumn{1}{l|}{}                                                                               & HQCNN-EM                   & \textbf{0.023 ± 0.007} & \textbf{0.881 ± 0.035} & \textbf{0.066 ± 0.023} & \textbf{0.556 ± 0.158} \\ \cline{2-7} 
\multicolumn{1}{c|}{}                       & \multicolumn{1}{l|}{\multirow{3}{*}{\begin{tabular}[c]{@{}l@{}}AerSim\\ (IBM-Yonsei)\end{tabular}}} & QCNN                       & 0.278 ± 0.016          & -0.424 ± 0.083         & 0.459 ± 0.074          & -2.110 ± 0.502         \\
\multicolumn{1}{c|}{}                       & \multicolumn{1}{l|}{}                                                                               & HQCNN-EZ                   & 0.031 ± 0.009          & 0.840 ± 0.047          & 0.098 ± 0.041          & 0.336 ± 0.280          \\
\multicolumn{1}{c|}{}                       & \multicolumn{1}{l|}{}                                                                               & HQCNN-EM                   & \textbf{0.022 ± 0.005} & \textbf{0.887 ± 0.024} & \textbf{0.076 ± 0.044} & \textbf{0.487 ± 0.298} \\ \midrule
\multicolumn{1}{c|}{\multirow{9}{*}{Test}}  & \multicolumn{1}{l|}{\multirow{3}{*}{noiseless}}                                                     & QCNN                       & 0.230 ± 0.018          & -0.725 ± 0.132         & 0.307 ± 0.070          & -2.065 ± 0.696         \\
\multicolumn{1}{c|}{}                       & \multicolumn{1}{l|}{}                                                                               & HQCNN-EZ                   & 0.017 ± 0.004          & 0.871 ± 0.031          & \textbf{0.053 ± 0.006} & \textbf{0.475 ± 0.063} \\
\multicolumn{1}{c|}{}                       & \multicolumn{1}{l|}{}                                                                               & HQCNN-EM                   & \textbf{0.014 ± 0.003} & \textbf{0.893 ± 0.018} & 0.055 ± 0.018          & 0.450 ± 0.176          \\ \cline{2-7} 
\multicolumn{1}{c|}{}                       & \multicolumn{1}{l|}{\multirow{3}{*}{FakeGuadalupeV2}}                                               & QCNN                       & 0.306 ± 0.015          & -1.289 ± 0.115         & 0.357 ± 0.076          & -2.567 ± 0.762         \\
\multicolumn{1}{c|}{}                       & \multicolumn{1}{l|}{}                                                                               & HQCNN-EZ                   & 0.026 ± 0.006          & 0.809 ± 0.048          & \textbf{0.062 ± 0.012} & \textbf{0.384 ± 0.122} \\
\multicolumn{1}{c|}{}                       & \multicolumn{1}{l|}{}                                                                               & HQCNN-EM                   & \textbf{0.013 ± 0.005} & \textbf{0.902 ± 0.035} & 0.064 ± 0.014          & 0.359 ± 0.143          \\ \cline{2-7} 
\multicolumn{1}{c|}{}                       & \multicolumn{1}{l|}{\multirow{3}{*}{\begin{tabular}[c]{@{}l@{}}AerSim\\ (IBM-Yonsei)\end{tabular}}} & QCNN                       & 0.322 ± 0.019          & -1.412 ± 0.143         & 0.401 ± 0.071          & -3.001 ± 0.712         \\
\multicolumn{1}{c|}{}                       & \multicolumn{1}{l|}{}                                                                               & HQCNN-EZ                   & 0.026 ± 0.006          & 0.806 ± 0.044          & \textbf{0.069 ± 0.025} & \textbf{0.310 ± 0.252} \\
\multicolumn{1}{c|}{}                       & \multicolumn{1}{l|}{}                                                                               & HQCNN-EM                   & \textbf{0.015 ± 0.003} & \textbf{0.886 ± 0.026} & 0.071 ± 0.018          & 0.289 ± 0.181  \\
\bottomrule
\end{tabular}
\caption{Training and test regression performance of QCNN and hybrid QCNN models across circuit sizes and noise settings. Results are reported for four- and eight-qubit circuits under noiseless and noisy simulations. Metrics include Mean Squared Error (MSE) loss and the coefficient of determination ($R^2$), reported as mean ± one standard deviation over five independent training runs. Arrows indicate the preferred direction of each metric. Bold values denote the best-performing model for each device and circuit size.}
\label{SI_tab:reg_performance}
\end{table}

\subsubsection*{Performance Scaling with Circuit Size}
\label{SI:subsec:reg_Performance_Advantage_w_circuit_size}
\begin{figure}[!t]
\centering
\includegraphics[width=0.98\textwidth]{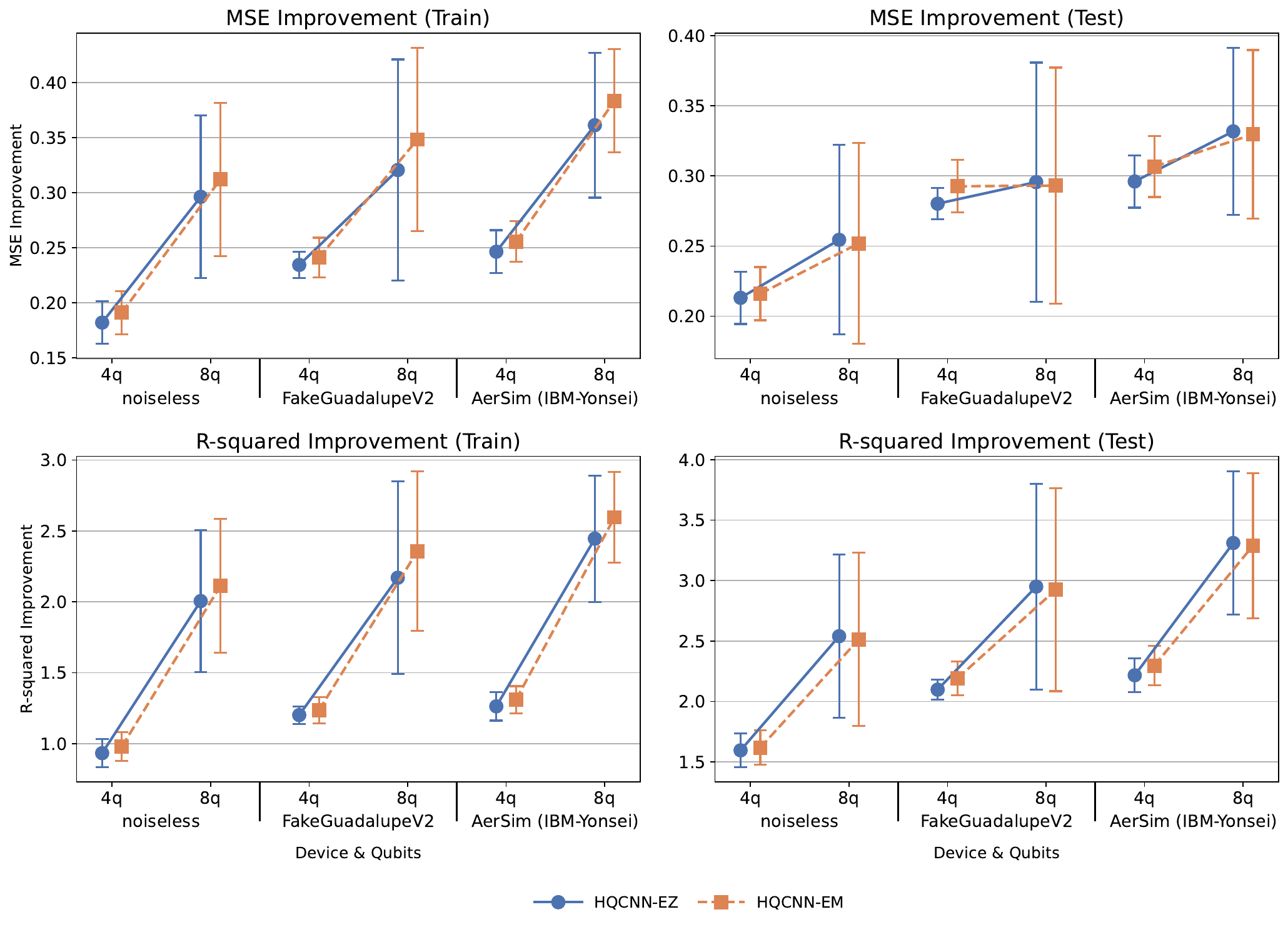}
\caption{Mean performance improvements of hybrid QCNN models relative to the baseline QCNN across circuit sizes and noise settings. Results are reported for the regression task using four- and eight-qubit circuits (denoted as \texttt{4q} and \texttt{8q}) under varying simulation conditions: noiseless, \texttt{FakeGuadalupeV2}, and \texttt{AerSimulator} configured with the \texttt{IBM\_Yonsei} backend. Metrics include the relative improvement in Mean Squared Error (MSE, top row) and Coefficient of Determination ($R^2$, bottom row), reported as mean $\pm$ one standard deviation over five independent training trials. Blue circles with solid lines denote HQCNN-EZ, while orange squares with dashed lines represent HQCNN-EM. Positive values of the relative improvement indicate a performance gain (i.e., lower MSE or higher $R^2$) over the baseline.}
\label{SI_fig:reg_perf_improve}
\end{figure}

To further explain the robustness of the proposed architecture, we examine how the performance advantage evolves as the circuit scale increases. Supplementary Figure~\ref{SI_fig:reg_perf_improve} visualizes the relative improvement of the hybrid models over the baseline QCNN across the tested noise environments, calculated using the same improvement metric formulated in the main text. As anticipated from the quantitative metrics in Supplementary Table~\ref{SI_tab:reg_performance}, both the HQCNN-EZ and HQCNN-EM architectures exhibit a highly similar trend of sustained improvement. This visual consistency reinforces the finding that the depth-stratified measurement strategy maintains a distinct predictive edge over the standard QCNN, effectively mitigating the performance degradation typically associated with increased circuit depth and noise accumulation.

\subsubsection*{SHAP analysis of depth-stratified measurements}
\label{SI:subsec:reg_SHAP_depth-stratified_meas}
\begin{figure}[!t]
\centering
\includegraphics[width=0.98\textwidth]{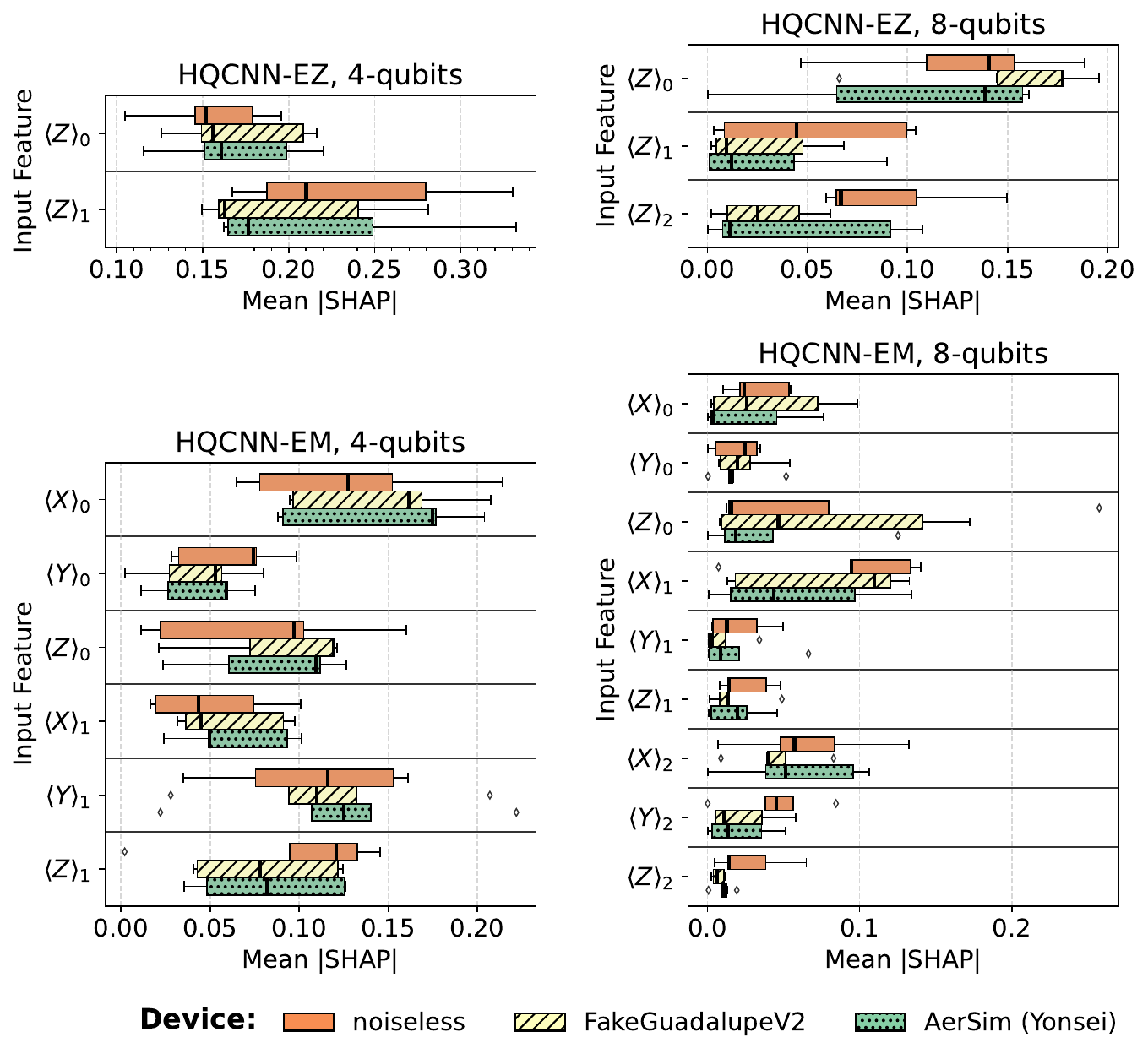}
\caption{Distribution of mean absolute SHAP values for depth-stratified measurements in the regression task across hybrid models, circuit sizes, and noise conditions. Global feature importance is quantified by the mean absolute SHAP value on the test dataset, where a higher magnitude indicates a stronger influence of the feature on the model's output. The box plots summarize the distribution of these scores across five independent trials, capturing stochastic variability. The central vertical line represents the median, serving as a robust indicator of importance. (Boxes: interquartile range; Whiskers: $1.5\times$ IQR; Points: outliers). Colors distinguish noise settings: noiseless (orange), \texttt{FakeGuadalupeV2} (yellow-hatched), and \texttt{AerSimulator} with \texttt{IBM\_Yonsei} backend (green-dotted). Features on the y-axis represent the expectation value $\langle P \rangle_i$, where $P$ denotes the measurement basis and $i$ the measured qubit index, ordered by circuit depth from shallow (top) to deep (bottom).}
\label{SI_fig:reg_abs_shap_boxplot}
\end{figure}

To confirm that the performance gains stem from the effective utilization of depth-stratified measurements, we analyzed the feature importance distributions using SHAP values. Supplementary Figure~\ref{SI_fig:reg_abs_shap_boxplot} presents the distribution of mean absolute SHAP values for the four- and eight-qubit hybrid models (HQCNN-EZ and HQCNN-EM) across all test samples.

Unlike the classification task, where feature importance was concentrated on specific early qubits, the regression task exhibits a more dynamic distribution of importance that varies with circuit depth. In the four-qubit regime, where the circuit depth is relatively shallow and noise accumulation is moderate, the HQCNN-EZ model assigns higher SHAP values to the deeper measurement ($\langle Z \rangle_1$) compared to the earlier one ($\langle Z \rangle_0$). This suggests that when noise is manageable, the model prefers the more transformed features from deeper layers. Similarly, the HQCNN-EM model in the four-qubit case displays a balanced distribution, where all Pauli measurements contribute significantly to the prediction, with $\langle X \rangle_0$ showing only a slight dominance.

However, a critical shift in behavior is observed in the eight-qubit models. As the circuit scales and noise accumulates, the HQCNN-EZ model shifts its focus upstream; the earliest measurement ($\langle Z \rangle_0$) becomes the most dominant feature, exhibiting significantly higher SHAP values than the deeper measurements ($\langle Z \rangle_1, \langle Z \rangle_2$). A similar trend is visible in the eight-qubit HQCNN-EM model, where early-to-intermediate features such as $\langle Z \rangle_0$ and $\langle X \rangle_1$ are identified as the most critical predictors. This adaptability demonstrates the architecture's noise resilience: as deeper quantum states degrade due to decoherence, the classical network learns to prioritize the higher-fidelity early-layer information to maintain predictive accuracy.

Overall, these results mirror the resilience observed in the classification task but highlight an added layer of adaptability. The architecture demonstrates the capacity to dynamically shift its reliance between deep and shallow features depending on the noise profile of the circuit, proving that the noise-mitigation properties are intrinsic to the design and effective across both discrete and continuous problem domains.

\clearpage

\subsection*{Extended training dynamics for classification task}
\label{SI:sec:clf_Extended_loss_convergence}
\begin{figure}[h]
    \centering
    \begin{subfigure}{0.9\textwidth}  
        \centering
        \includegraphics[width=\linewidth]{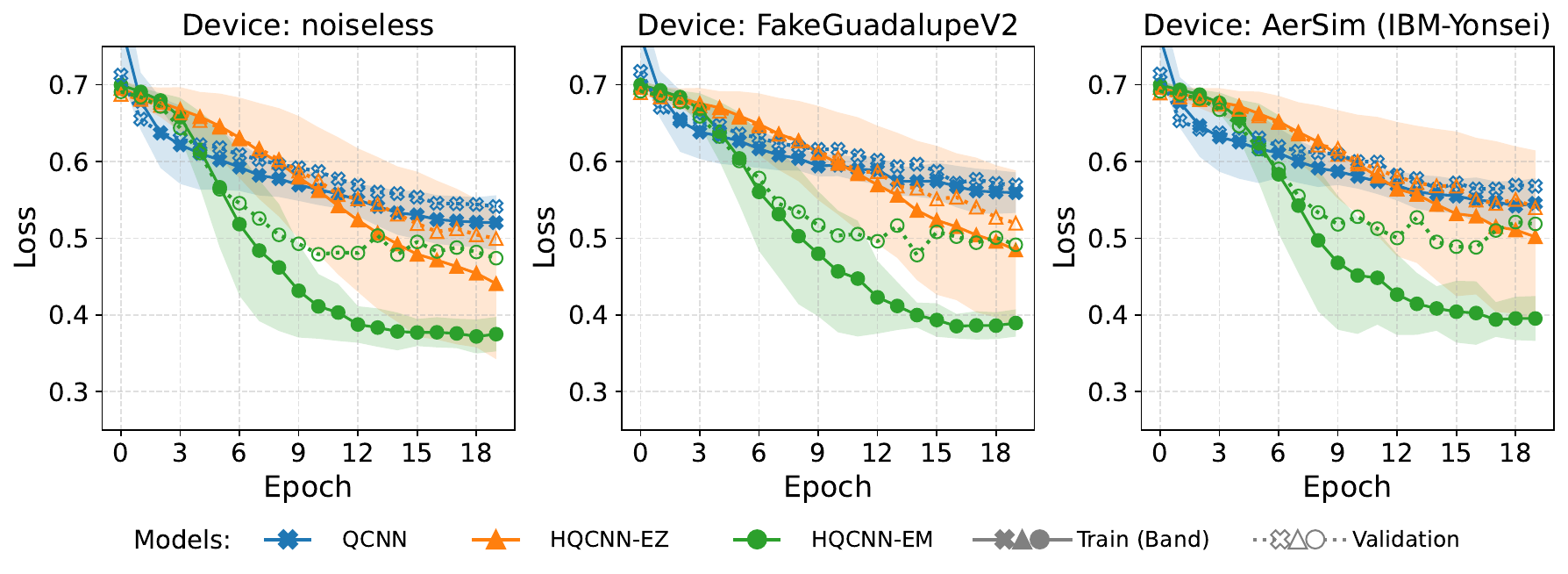}
        \caption{Training loss of four-qubit circuit}
        \label{SI_fig:clf_loss_convergence_q4}
    \end{subfigure}

    \vspace{0.8em}  


    \begin{subfigure}{0.9\textwidth}
        \centering
        \includegraphics[width=\linewidth]{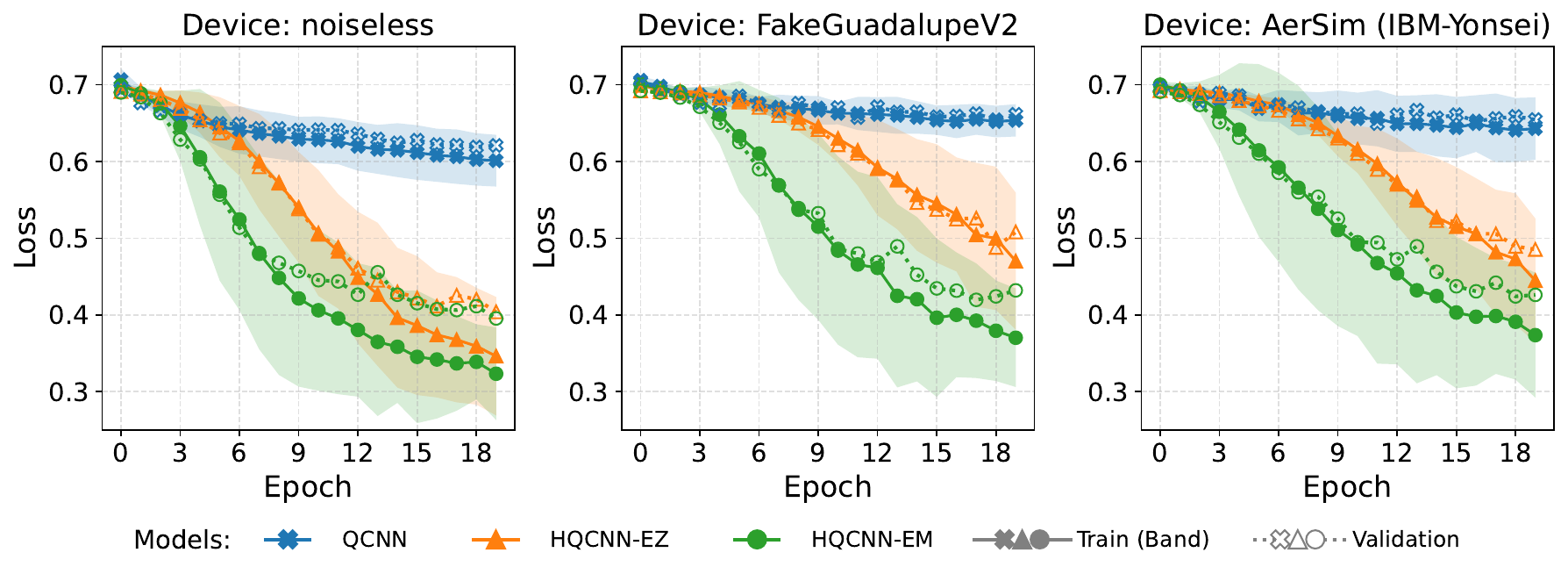}
        \caption{Training loss of ten-qubit circuit}
        \label{SI_fig:clf_loss_convergence_q10}
    \end{subfigure}
    
    \caption{Training and validation loss curves of the classification models under varying simulation conditions. The plots display the loss trajectories for (a) a four-qubit circuit and (b) a ten-qubit circuit across three environments: noiseless, \texttt{FakeGuadalupeV2}, and \texttt{AerSimulator} configured with the \texttt{IBM\_Yonsei} backend. Solid and dotted lines represent the mean of training and validation loss, respectively, averaged over five independent iterations. The shaded bands around the training loss curves indicate one standard deviation ($\pm \sigma$) around the mean.}
    \label{SI_fig:clf_loss_convergence}
\end{figure}

To verify the consistency of the training stability observed in the main text, we examine the loss trajectories for four-qubit and ten-qubit classification models. Supplementary Figure~\ref{SI_fig:clf_loss_convergence} presents the training and validation BCE loss curves across noiseless and noisy simulation environments.

In the four-qubit regime (Supplementary Figure~\ref{SI_fig:clf_loss_convergence_q4}), the training dynamics reveal distinct behaviors among the models. Initially, the baseline QCNN and the HQCNN-EZ model exhibit similar loss trajectories during the first half of the training epochs. However, in the latter half, the HQCNN-EZ model achieves a further reduction in loss, gradually diverging from the baseline, which stabilizes at a higher value. Notably, as training progresses, the HQCNN-EM model demonstrates an improved loss trajectory, exhibiting a steeper descent and converging to the lowest final BCE loss values among all tested models.

The enhanced trainability of the hybrid architectures becomes significantly more pronounced in the ten-qubit regime (Supplementary Figure~\ref{SI_fig:clf_loss_convergence_q10}). Here, the baseline QCNN exhibits signs of barren plateaus, characterized by a stagnant loss curve that fails to improve beyond the initial state. In sharp contrast, both hybrid models circumvent this training limitation, maintaining a steep learning trajectory and converging to low loss values even in the presence of realistic hardware noise. This consistency confirms that the depth-stratified measurement strategy effectively mitigates the trainability issues that typically impede deeper quantum classifiers.

\subsection*{Calibration Data}
To quantify the noise conditions used in the simulations, we recorded calibration metrics for the physical qubits (including ancilla qubits) assigned to each transpiled quantum circuit. Supplementary Tables~\ref{SI_tab:clf_backend_calibration} and \ref{SI_tab:reg_backend_calibration} provide the aggregated mean and standard deviation of these hardware metrics across all devices and models for each classification and regression task, respectively. The recorded calibration metrics are coherence times ($T_1$, $T_2$), qubit frequencies, and anharmonicity; readout properties (assignment error, SPAM probabilities, and readout length); single-qubit gate errors (ID, single-qubit gate length, $RZ$, $\sqrt{X}$, $X$) and durations; and finally two-qubit gate errors (CNOT/ECR) and gate duration; and measurement operation error.

\begin{landscape}
\begin{table}[p]
\centering
\footnotesize 
\setlength{\tabcolsep}{3pt} 
\begin{tabular}{l|l|c|c|c|c|c}
\toprule
\multicolumn{1}{c|}{\textbf{Device}} & \multicolumn{1}{c|}{\textbf{Model}} & \textbf{T1 ($\mu$s)} & \textbf{T2 ($\mu$s)} & \textbf{Frequency (GHz)} & \textbf{Anharmonicity (GHz)} & \textbf{Readout assignment error} \\ \hline
\multirow{11}{*}{\begin{tabular}[c]{@{}l@{}}Fake\\ GuadalupeV2\end{tabular}} & QCNN & 74.779 ± 4.396 & 93.662 ± 5.487 & 5.209 ± 0.041 & -0.326 ± $8.053\times10^{-4}$ & 0.0174 ± 0.00256 \\
 & HQCNN-EZ & 74.205 ± 5.007 & 93.914 ± 4.764 & 5.215 ± 0.045 & -0.326 ± $1.011\times10^{-3}$ & 0.0183 ± 0.00325 \\
 & HQCNN-EM & 74.472 ± 4.971 & 94.149 ± 4.561 & 5.217 ± 0.048 & -0.326 ± $1.087\times10^{-3}$ & 0.0187 ± 0.00344 \\ \cmidrule{2-7} 
 & \multicolumn{1}{c|}{\textbf{}} & \textbf{Prob meas0 prep1} & \textbf{Prob meas1 prep0} & \textbf{Readout length (ns)} & \textbf{ID error} & \textbf{Single-qubit gate length (ns)} \\ \cline{2-7} 
 & QCNN & 0.0288 ± 0.00345 & 0.0060 ± 0.00188 & 5351.11 & $3.312\times10^{-4}$ ± $7.947\times10^{-5}$ & 35.556 ± 0.0 \\
 & HQCNN-EZ & 0.0300 ± 0.00441 & 0.0067 ± 0.00230 & 5351.11 & $3.639\times10^{-4}$ ± $9.464\times10^{-5}$ & 35.556 ± 0.0 \\
 & HQCNN-EM & 0.0306 ± 0.00468 & 0.0068 ± 0.00239 & 5351.11 & $3.639\times10^{-4}$ ± $9.427\times10^{-5}$ & 35.556 ± 0.0 \\ \cmidrule{2-7} 
 & \multicolumn{1}{c|}{\textbf{}} & \textbf{$\sqrt{\text{x}}$ (sx) error} & \textbf{Pauli-X error} & \textbf{CNOT/ECR error} & \textbf{Gate length (ns)} & \textbf{MEASURE error} \\ \cline{2-7} 
 & QCNN & $3.312\times10^{-4}$ ± $7.950\times10^{-5}$ & $3.312\times10^{-4}$ ± $7.950\times10^{-5}$ & 0.0103 ± $3.823\times10^{-4}$ & 386.908 ± 9.930 & 0.017 ± 0.003 \\
 & HQCNN-EZ & $3.639\times10^{-4}$ ± $9.460\times10^{-5}$ & $3.639\times10^{-4}$ ± $9.460\times10^{-5}$ & 0.0104 ± $4.226\times10^{-4}$ & 390.255 ± 10.805 & 0.018 ± 0.003 \\
 & HQCNN-EM & $3.639\times10^{-4}$ ± $9.430\times10^{-5}$ & $3.639\times10^{-4}$ ± $9.430\times10^{-5}$ & 0.0105 ± $4.882\times10^{-4}$ & 391.664 ± 9.481 & 0.019 ± 0.003 \\ \cmidrule{1-7}
\textbf{} & \multicolumn{1}{c|}{} & \textbf{T1 ($\mu$s)} & \textbf{T2 ($\mu$s)} & \textbf{Frequency (GHz)} & \textbf{Anharmonicity (GHz)} & \textbf{Readout assignment error} \\ \cline{2-7}
\multirow{11}{*}{\begin{tabular}[c]{@{}l@{}}AerSim\\ (IBM-Yonsei)\end{tabular}} & QCNN & 267.589 ± 20.595 & 152.953 ± 33.559 & 4.839 ± 0.043 & -0.308 ± $1.016\times10^{-3}$ & 0.0203 ± 0.01063 \\
 & HQCNN-EZ & 272.905 ± 24.541 & 155.928 ± 36.475 & 4.831 ± 0.051 & -0.309 ± $1.350\times10^{-3}$ & 0.0224 ± 0.01161 \\
 & HQCNN-EM & 266.791 ± 17.863 & 159.786 ± 38.836 & 4.829 ± 0.030 & -0.309 ± $4.404\times10^{-4}$ & 0.0159 ± 0.00308 \\ \cmidrule{2-7} 
 &  & \textbf{Prob meas0 prep1} & \textbf{Prob meas1 prep0} & \textbf{Readout length (ns)} & \textbf{ID error} & \textbf{Single-qubit gate length (ns)} \\ \cline{2-7} 
 & QCNN & 0.0218 ± 0.01057 & 0.0171 ± 0.00934 & 840.00 & $3.009\times10^{-4}$ ± $1.107\times10^{-4}$ & 60.000 ± 0.0 \\
 & HQCNN-EZ & 0.0218 ± 0.00799 & 0.0162 ± 0.00813 & 840.00 & $3.399\times10^{-4}$ ± $2.202\times10^{-4}$ & 60.000 ± 0.0 \\
 & HQCNN-EM & 0.0182 ± 0.00435 & 0.0126 ± 0.00411 & 840.00 & $2.653\times10^{-4}$ ± $6.413\times10^{-5}$ & 60.000 ± 0.0 \\ \cmidrule{2-7}
 &  & \textbf{$\sqrt{\text{x}}$ (sx) error} & \textbf{Pauli-X error} & \textbf{CNOT/ECR error} & \textbf{Gate length (ns)} & \textbf{MEASURE error} \\ \cline{2-7} 
 & QCNN & $3.009\times10^{-4}$ ± $1.107\times10^{-4}$ & $3.009\times10^{-4}$ ± $1.107\times10^{-4}$ & 0.0068 ± $8.536\times10^{-4}$ & 638.953 ± 3.564 & 0.020 ± 0.011 \\
 & HQCNN-EZ & $3.399\times10^{-4}$ ± $2.202\times10^{-4}$ & $3.399\times10^{-4}$ ± $2.202\times10^{-4}$ & 0.0090 ± $5.570\times10^{-3}$ & 638.578 ± 2.643 & 0.022 ± 0.012 \\
 & HQCNN-EM & $2.653\times10^{-4}$ ± $6.410\times10^{-5}$ & $2.653\times10^{-4}$ ± $6.410\times10^{-5}$ & 0.0070 ± $5.335\times10^{-4}$ & 639.480 ± 3.041 & 0.016 ± 0.003 \\
\bottomrule
\end{tabular}
\caption{Aggregated backend calibration metrics for the classification task}
\label{SI_tab:clf_backend_calibration}
\end{table}
\end{landscape}

\begin{landscape}
\begin{table}[p]
\centering
\footnotesize 
\setlength{\tabcolsep}{3pt} 
\begin{tabular}{l|l|c|c|c|c|c}
\toprule
\multicolumn{1}{c|}{\textbf{Device}} & \multicolumn{1}{c|}{\textbf{Model}} & \textbf{T1 ($\mu$s)} & \textbf{T2 ($\mu$s)} & \textbf{Frequency (GHz)} & \textbf{Anharmonicity (GHz)} & \textbf{Readout assignment error} \\ \hline
\multirow{11}{*}{\begin{tabular}[c]{@{}l@{}}Fake\\ GuadalupeV2\end{tabular}} & QCNN & 74.816 ± 4.667 & 94.076 ± 4.627 & 5.216 ± 0.046 & -0.326 ± $9.462\times10^{-4}$ & 0.0187 ± 0.00339 \\
 & HQCNN-EZ & 74.354 ± 4.446 & 93.117 ± 5.215 & 5.216 ± 0.045 & -0.326 ± $9.451\times10^{-4}$ & 0.0183 ± 0.00340 \\
 & HQCNN-EM & 74.417 ± 4.550 & 94.339 ± 5.140 & 5.213 ± 0.043 & -0.326 ± $9.567\times10^{-4}$ & 0.0184 ± 0.00305 \\ \cmidrule{2-7}
 & \multicolumn{1}{c|}{\textbf{}} & \textbf{Prob meas0 prep1} & \textbf{Prob meas1 prep0} & \textbf{Readout length (ns)} & \textbf{ID error} & \textbf{Single-qubit gate length (ns)} \\ \cline{2-7} 
 & QCNN & 0.0307 ± 0.00458 & 0.0068 ± 0.00237 & 5351.11 & $3.647\times10^{-4}$ ± $9.484\times10^{-5}$ & 35.556 ± 0.0 \\
 & HQCNN-EZ & 0.0298 ± 0.00473 & 0.0067 ± 0.00231 & 5351.11 & $3.779\times10^{-4}$ ± $9.719\times10^{-5}$ & 35.556 ± 0.0 \\
 & HQCNN-EM & 0.0303 ± 0.00415 & 0.0065 ± 0.00214 & 5351.11 & $3.558\times10^{-4}$ ± $9.279\times10^{-5}$ & 35.556 ± 0.0 \\ \cmidrule{2-7}
 & \multicolumn{1}{c|}{\textbf{}} & \textbf{$\sqrt{x}$ (sx) error} & \textbf{Pauli-X error} & \textbf{CNOT/ECR error} & \textbf{Gate length (ns)} & \textbf{MEASURE error} \\ \cline{2-7} 
 & QCNN & $3.647\times10^{-4}$ ± $9.480\times10^{-5}$ & $3.647\times10^{-4}$ ± $9.480\times10^{-5}$ & 0.0105 ± $4.816\times10^{-4}$ & 393.007 ± 10.300 & 0.019 ± 0.003 \\
 & HQCNN-EZ & $3.779\times10^{-4}$ ± $9.720\times10^{-5}$ & $3.779\times10^{-4}$ ± $9.720\times10^{-5}$ & 0.0105 ± $4.561\times10^{-4}$ & 392.728 ± 10.903 & 0.018 ± 0.003 \\
 & HQCNN-EM & $3.558\times10^{-4}$ ± $9.280\times10^{-5}$ & $3.558\times10^{-4}$ ± $9.280\times10^{-5}$ & 0.0104 ± $4.029\times10^{-4}$ & 390.685 ± 10.538 & 0.018 ± 0.003 \\ \cmidrule{1-7}
\textbf{} & \multicolumn{1}{c|}{} & \textbf{T1 ($\mu$s)} & \textbf{T2 ($\mu$s)} & \textbf{Frequency (GHz)} & \textbf{Anharmonicity (GHz)} & \textbf{Readout assignment error} \\ \cline{2-7}
\multirow{11}{*}{\begin{tabular}[c]{@{}l@{}}AerSim\\ (IBM-Yonsei)\end{tabular}} & QCNN & 257.255 ± 23.033 & 145.562 ± 73.335 & 4.816 ± 0.030 & -0.309 ± $9.989\times10^{-4}$ & 0.0208 ± 0.01132 \\
 & HQCNN-EZ & 263.576 ± 15.613 & 176.960 ± 17.442 & 4.786 ± 0.047 & -0.310 ± $1.194\times10^{-3}$ & 0.0145 ± 0.00663 \\
 & HQCNN-EM & 259.506 ± 11.621 & 182.167 ± 30.769 & 4.786 ± 0.033 & -0.310 ± $1.056\times10^{-3}$ & 0.0148 ± 0.00874 \\ \cmidrule{2-7}
 &  & \textbf{Prob meas0 prep1} & \textbf{Prob meas1 prep0} & \textbf{Readout length (ns)} & \textbf{ID error} & \textbf{Single-qubit gate length (ns)} \\ \cline{2-7} 
 & QCNN & 0.0219 ± 0.01048 & 0.0166 ± 0.00502 & 840.00 & $3.635\times10^{-4}$ ± $1.932\times10^{-4}$ & 60.000 ± 0.0 \\
 & HQCNN-EZ & 0.0165 ± 0.00564 & 0.0135 ± 0.00776 & 840.00 & $2.698\times10^{-4}$ ± $4.277\times10^{-5}$ & 60.000 ± 0.0 \\
 & HQCNN-EM & 0.0171 ± 0.00826 & 0.0132 ± 0.00911 & 840.00 & $2.898\times10^{-4}$ ± $5.034\times10^{-5}$ & 60.000 ± 0.0 \\ \cmidrule{2-7}
 &  & \textbf{$\sqrt{\text{x}}$ (sx) error} & \textbf{Pauli-X error} & \textbf{CNOT/ECR error} & \textbf{Gate length (ns)} & \textbf{MEASURE error} \\ \cline{2-7} 
 & QCNN & $3.635\times10^{-4}$ ± $1.932\times10^{-4}$ & $3.635\times10^{-4}$ ± $1.932\times10^{-4}$ & 0.0086 ± $4.983\times10^{-3}$ & 638.413 ± 2.670 & 0.021 ± 0.011 \\
 & HQCNN-EZ & $2.698\times10^{-4}$ ± $4.280\times10^{-5}$ & $2.698\times10^{-4}$ ± $4.280\times10^{-5}$ & 0.0066 ± $8.335\times10^{-4}$ & 640.537 ± 2.038 & 0.014 ± 0.007 \\
 & HQCNN-EM & $2.898\times10^{-4}$ ± $5.030\times10^{-5}$ & $2.898\times10^{-4}$ ± $5.030\times10^{-5}$ & 0.0067 ± $9.718\times10^{-4}$ & 639.856 ± 2.131 & 0.015 ± 0.009 \\
\bottomrule
\end{tabular}
\caption{Aggregated backend calibration metrics for the regression task}
\label{SI_tab:reg_backend_calibration}
\end{table}
\end{landscape}

\end{document}